%% file: main.tex
\documentclass[10pt, conference]{IEEEtran}

\usepackage[english]{babel}
\usepackage[utf8]{inputenc}
\usepackage{amsmath}
\usepackage{amssymb}
\usepackage{graphicx}
\usepackage{cite}
\usepackage{url}
\usepackage{pgfplots}
\usepackage{multirow}
\usepackage{tabularx}
\usepackage{booktabs}
\usepackage{makecell}
\usepackage{float}
\usepackage{array}
\usepackage{listings}
\usepackage{lipsum}
\usepackage[most]{tcolorbox}
\usepackage{algorithm}
\usepackage{algpseudocode}
\usepackage{subcaption}
\usepackage{listings}
\usepackage[style=base]{caption}
\usepackage[T1]{fontenc} 
\usepackage{adjustbox}
\usepackage{graphicx} 
\usepackage{tabularx} 
\usepackage{bm} 

\usepackage{color}
\usepackage{colortbl}

\newcolumntype{L}[1]{>{\raggedright\arraybackslash}m{#1}}
\newcolumntype{C}[1]{>{\centering\arraybackslash}m{#1}}
\newcolumntype{R}[1]{>{\raggedleft\arraybackslash}m{#1}}

\algnewcommand\algorithmicforeach{\textbf{for each}}
\algdef{S}[FOR]{ForEach}[1]{\algorithmicforeach\ #1\ \algorithmicdo}
\algdef{SE}[DOWHILE]{Do}{doWhile}{\algorithmicdo}[1]{\algorithmicwhile\ #1}%

\newcommand{\ie}{\emph{\mbox{i.e.,}}}

\definecolor{primary}{RGB}{46,125,50}

\newcommand\customcomment[1]{{\color{blue} \sffamily [xxx:  #1]}}
\newcommand\reviewfix[1]{{\color{green}\sffamily [RF:#1]}}   
\newcommand\PostSubmission[1]{{\sffamily [\textcolor{blue}{post-xxx:  #1]}}}
\makeatother

\def\Snospace~{\S{}}
\newcommand{\dns}[1]{{\small \texttt{#1}}}

\title{Hosting Industry Centralization and Consolidation}
\author{
  \IEEEauthorblockN{Luciano Zembruzki\IEEEauthorrefmark{1}, Raffaele Sommese\IEEEauthorrefmark{2}, Lisandro Zambenedetti Granville\IEEEauthorrefmark{1},\\Arthur Selle Jacobs\IEEEauthorrefmark{1}, Mattijs Jonker\IEEEauthorrefmark{2}, Giovane C. M.  Moura\IEEEauthorrefmark{3}}\\
  \IEEEauthorblockA{
        \begin{tabular}{ccc}
            \begin{tabular}{@{}c@{}}
                \IEEEauthorrefmark{1}Institute of Informatics (INF) -\\
                    Federal University of Rio\\Grande do Sul (UFRGS)\\
                    \textit{Porto Alegre, Brazil}\\
                    \{lzembruzki, asjacobs, granville\}@inf.ufrgs.br
            \end{tabular} & \begin{tabular}{@{}c@{}}
                \IEEEauthorrefmark{2}Design and Analysis of\\Communication Systems (DACS) - \\
                    University of Twente\\
                    \textit{Enschede, The Netherlands}\\
                    \{r.sommese, m.jonker\}@utwente.nl
            \end{tabular} & \begin{tabular}{@{}c@{}}
                \IEEEauthorrefmark{3}SIDN\\
                    SIDN Labs\\
                    \textit{Arnhem, The Netherlands}\\
                    giovane.moura@sidn.nl
            \end{tabular}
        \end{tabular} 
  }
}

\IEEEoverridecommandlockouts \IEEEpubid{\makebox[\columnwidth]{978-1-6654-0601-7/22\/\$31.00 ~\copyright~2022 IEEE \hfill} \hspace{\columnsep}\makebox[\columnwidth]{ } }

\begin{document}
\maketitle

\input{src/abstract}
\input{src/introduction}
\input{src/related_work}
\input{src/methodology.tex}

\input{src/hosting_concentration}

\input{src/concentration_per_TLD}
\input{src/conclusion}
\input{src/acknowledgements}

\bibliographystyle{IEEEtran}
\bibliography{IEEEabrv,bibtex,paper,rfc}{}

\end{document}

%% file: src/abstract.tex
\begin{abstract}
There have been growing concerns about the concentration and centralization of Internet infrastructure. In this work, we scrutinize the hosting industry on the Internet by using active measurements, covering 19 Top-Level Domains~(TLDs). We show how the market is heavily concentrated: 1/3 of the domains are hosted by only 5 hosting providers, all US-based companies. For  the country-code TLDs~(ccTLDs), however, hosting is primarily done by local, national hosting providers and not by the large American cloud and content providers. We show how shared languages (and borders) shape the hosting market --- German hosting companies have a notable presence in Austrian and Swiss markets, given they all share German as official language. While hosting concentration has been relatively high and stable over the past four years, we see that American hosting companies have been continuously increasing their presence in the market related to high traffic, popular domains within ccTLDs --- except for Russia, notably.

\end{abstract}

%% file: src/introduction.tex
\section{Introduction}
\label{sec:intro}

Internet centralization and consolidation refers to the concentration of, \textit{e.g.}, user base, infrastructure, and network traffic in the hands of few, yet large Internet market players. Centralization raises concerns among diverse entities and individuals such as academics, operators, not-for-profit organizations, standardization bodies, the European Commission, the US Department of Justice, policy makers, and civil society \cite{Arkko2020a,arkko-draft-ietf-centralization,draft-arkko-iab-internet-consolidation-02,Schneier2018a,isoc19a,Kang20a,Moura20a,Kashaf20a,McCabe20a,Satariano20a,zuboff2019age,Wu2018a,Erlanger20a}. Internet consolidation usually implies concentration of power too \cite{zuboff2019age} \cite{Wu2018a} -- including political power, as seen in the last US presidential election \cite{Roose21a} \cite{Nicas21a}. From a technical perspective, concentration in the hands of few market players can also create large, single points of failure. Notable examples of this include Denial-of-Service (DoS) attacks against the DNS providers Dyn and AWS, which affected multiple popular websites and services \cite{Perlroth16a} \cite{aws-ddos}, including Twitter, Netflix, and Spotify. Consolidation can also lead to a large attack surface in which ``bad decisions or poorly made trade-offs implemented by a company can quickly scale to hundreds of millions of users'' \cite{Schneier21a}, as in the case of the Sunburst cyber espionage campaign where Microsoft, a large cloud provider, was hacked.

Quantifying Internet centralization is a challenging task that encompasses questions such as: what exactly to measure and where? Recent studies have either measured \textit{traffic} as a way to assess centralization (\textit{e.g.}, Moura \textit{et al.} \cite{Moura20a} found that 1/3 of the DNS traffic towards the country-code Top-Level Domains (ccTLDs) of the Netherlands and New Zealand originated from 5 large, American companies) or quantified centralization in terms of \textit{infrastructure} concentration (Kashaf \textit{et al.} \cite{Kashaf20a} and Allman \cite{Allman18a} studied authoritative DNS \cite{rfc8499} service infrastructure concentration). Centralization can also be quantified in terms of \textit{user base} or \textit{market share}. Facebook, for example, has 2.9B monthly active users (Jul. 2021 \cite{Facebook-QR2021}). In general, all these consolidation assessments complement one another and highlight different aspects of Internet centralization.



In contrast with the aforementioned efforts, in this work we focus on analyzing consolidation in the \textit{Web hosting} industry~\cite{5353074}, which is a market segment dedicated to hosting websites and services that use HTTP/S \cite{RFC7540} as application protocol. In 2020, this market segment was valued at US\$ 75B \cite{Fortune20a}. Yet, the hosting industry has been mostly neglected in state-of-the-art Internet centralization studies. To close this gap, we develop a methodology that employs DNS zone files and active measurements to map \textit{all} domains from 19 Top-Level domains (TLDs), including sixteen ccTLDs and three generic TLDs (gTLDs), to their respective Web hosting companies. In addition, we analyze current and historical measurement data, spanning over five years, from 2017 to 2021, to observe the development (and increase) of Web hosting centralization. 

The contribution of this paper is threefold. First, we reveal that \textit{Web hosting} is heavily concentrated by finding that: \textit{(i)} more than 1/3 of 150 million analyzed domains from 19 TLDs are hosted by five large US hosting providers;  \textit{(ii)} as a result of \textit{one policy} change by CloudFlare (the largest 2020 hosting provider in our dataset), $\sim$17\,M domains were moved to Google Cloud, increasing the concentration of domain names and turning Google into the largest provider in 2021, with 18\% of all domains; and \textit{(iii)} although the centralization percentage varies by each TLD, most of them concentrate at least 40\% of domains in five hosting providers. In addition, this centralization has been increasing over the past five years. As second contribution, we show that geographical proximity and shared language ties influence the hosting industry; although European ccTLDs have a solid local hosting industry, German-speaking countries in Europe (Austria, Germany, Switzerland, and Liechtenstein) use each other's hosting provider industry. Canada \dns{.ca}, in turn, mainly relies on US-based hosting companies. Our third and final contribution is to show that US-based hosting companies have been increasing their market share on popular domains for most TLDs, which poses a challenge for European Union (EU) digital sovereignty goals \cite{Erlanger20a}. The exception is Russia's ccTLDs (\dns{.ru} and \dns{.$p\phi$}), for which most popular domains are hosted by local companies.

The remainder of this paper is organized as follows. In Section II, we present related work on measurements of different aspects of Internet centralization and consolidation, in addition of positioning our work with regard to the state-of-the-art. In Section III, we describe our analysis methodology, limitations and data sets obtained from the OpenINTEL project. Then, in Sections IV and V, we present results and show the market share and top DNS providers by year, TLD, and geographical location. Finally, in Section VI, we draw conclusions and outline future work.

%% file: src/related_work.tex
\section{Related Work}
\label{sec:relatedwork}

The centralization and consolidation of the Internet is a known issue. Several efforts have been made to quantify to which extent this is affecting the global hosting ecosystem. However, as maintainers and DNS providers infrequently disclose DNS zone files and rarely share their internal configurations (aggregated or otherwise), measuring concentration is daunting. To that end, authors have taken different approaches to circumvent the general lack of transparency, often relying on active measurement techniques. In addition, identifying industry trends over time requires long-term data collection efforts, which come with a number of infrastructural challenges for researchers. In this section, we first briefly survey the state-of-the-art on DNS measurements, and then visit past efforts to quantify centralization and consolidation of hosting infrastructures.

Bates \textit{et al.} \cite{bates2018evidence} proposed a method to compute to which extent the global DNS infrastructure has maintained its distributed resilience in the face of the development of cloud-based DNS infrastructures. The authors examined the concentration and diversity of DNS over time while considering only \dns{.com}, \dns{.net}, and \dns{.org} domain names on the list of Alexa Top 1k domains. The authors argued that they limited their investigation to the TLDs \dns{.com}, \dns{.net}, and \dns{.org} because these are among the oldest and represent a broad part of the Internet. The authors, however, acknowledge that their findings may vary if additional TLDs, such as \dns{.ru} and \dns{.cn}, would be be taken into account.

Kashaf \textit{et al.} \cite{Kashaf20a} studied the presence and impact of third-party dependencies in three infrastructure services: DNS, CDNs, and certificate revocation checks by CAs. They analyzed both direct and indirect dependencies. The authors found that 89\% of Alexa's Top 100\,k websites rely crucially on third-party DNS, CDN, or CA providers, implying that these websites will experience service disruption if these third-party providers fail. The study also shows that third-party service use is concentrated, with the leading Top 3 providers of CDN, DNS, or CA services affecting between 50\% to 70\% of the Top 100\,k websites. It is worth mentioning that, in their analysis of DNS dependencies, the authors rely solely on NS record labels (\textit{e.g.,} \dns{ns1.example.com}) to measure concentration. Since multiple name servers can be hosted on a single IP address, the authors' analysis can mask possible cases of centralization.

Moura \textit{et al.}~\cite{Moura20a} Using an investigation of DNS traffic acquired at a DNS root server and two ccTLDs (one in Europe and one in Oceania),  discovered indicators of concentration: over 30\% of all requests to both ccTLDs were made via five major cloud providers (Google, Amazon, Microsoft, Facebook, and Cloudflare). Unlike earlier initiatives, the authors underlined one benefit of centralization: whenever the cloud provider improves its infrastructure, for example, in terms of security features like QNAME reduction, a large number of clients benefit immediately.

The aforementioned research efforts inspect and measure different aspects of infrastructure to quantify the centralization and consolidation of DNS. However, they do not consider centralization of the hosting industry, which plays a key role in the Internet ecosystem as well. Although, from a technical perspective, analyzing the DNS infrastructures may provide a general view of the problem space, it is by analyzing the concentration of domains per hosting provider that one can shed light onto the often overlooked issue of hosting industry centralization.

Tajalizadehkhoob \textit{et al.} \cite{tajalizadehkhoob2016apples} presented a method to capture the hosting market's complexity. They identify and classify hosting providers by using a technique that combines: (i) passive DNS data to locate hosting infrastructure, and (ii) WHOIS data to address that infrastructure. In doing so, they are able to quantify providers and look at geographic distribution. The authors further exploit features to characterize the hosting provider landscape and reduce hosting market complexity and heterogeneity by using cluster analysis. They discovered 45,434 hosting providers dispersed over 1,517 IP addresses on average. Although the hosting services are commoditized -- and thus susceptible to economies of scale, as witnessed in the cloud services sector -- the authors surprisingly found little consolidation in the market.

Ager \textit{et al.} \cite{ager2011web} proposed an approach to infer and categorize hosting infrastructures. The authors rely on data available to end-users that request hostnames via the DNS and further use information such as IP addresses, prefixes, and AS numbers to create a mapping of hosting infrastructures. Their findings demonstrate that qualitative observations can be made for the establishment of hosting infrastructure and content replication. One of the study's primary findings is that a considerable portion of content is given solely by hosting infrastructures such as Google, or in geographic locations such as China, the latter of which suggests that spoken language could be a factor.

To the best of our knowledge, no prior work has provided a comprehensive overview nor comparison of centralization and consolidation by ccTLDs and gTLDs. Furthermore, the relation between spoken language, hosting provider location, and market division, as we do in this paper, was not previously investigated.

%% file: src/methodology.tex
\section{Methodological Overview}

In this section, we detail the methodology and data sources that we used to analyze both the current landscape of the hosting industry as well as its historical evolution. We also motivate assumptions made and discuss methodological limitations. Finally, we describe the concrete data set used for this work.

\subsection{Methodology}
\label{sec:methodology}


We rely on large-scale DNS measurement data for analysis. Data are provided by the OpenINTEL project \cite{van16}, which measures roughly 65\% of the global DNS namespace (second-level domains) on a daily basis.  OpenINTEL primarily takes a set of zone files as seed and enumerates, through active querying, the resource records (RRs) configured under domain names. OpenINTEL covers: generic top-level domains~(gTLDs) such as \dns{.com}, \dns{.net}, and \dns{.org}; country-code zones (ccTLDs) for several continents (\textit{e.g.}, \dns{.ca}); and about 1,200 new gTLDs such as \dns{.tokyo}\footnote{Zone file access typically requires specific agreements with registries that operate under varying regulations, which means that covering every possible top-level domain, especially some ccTLDs, is unfeasible.}. In addition to zone files, a number of other sources of domains are measured (\textit{e.g.}, the Alexa Top 1\,M).

We use other data sources to enrich or complement the OpenINTEL DNS data. For example, we use the Public Suffix List~\cite{web:psl}, a community-driven initiative of Mozilla, to demarcate top-level domains (\textit{e.g.}, \dns{.co.uk}). We use daily CAIDA prefix-to-AS data~\cite{caida_pfx2as} to map IP addresses (\ie~\texttt{A} and \texttt{AAAA} records) to the announcing AS number. We leverage monthly AS-to-organization data~\cite{caida-as2org-data} to map AS numbers to organizations and countries. Finally, we use an open countries and languages data set to map country codes to (one or more) official/predominant languages~\cite{web:annexare:countries}. It is important to note that this mapping is different from IP geolocation~\cite{muir_internet_2009,Poese:2011:IGD:1971162.1971171}, which maps IP addresses to geographical locations. As we will show later on, we are instead interested in the country to which the announcing AS can be associated.

\subsection{Assumptions and Limitations}

Our methodology involves a number of assumptions and also a few limitations:

\textbf{Web hosting.} For simplicity, we assume that the presence of an \texttt{A} or \texttt{AAAA} record for a domain name (or its \dns{www.} label) signals Web hosting. This assumption will break for domain names strictly used for non Web purposes, but we expect this to have limited impact on our results. As we will show in section \ref{sec:results}, the vast majority of domain names configure the \texttt{www} label, which is a reasonable indicator of Web hosting intent. We also checked this assumption in May 2021 HTTP crawl data of \texttt{.nl} and found that: 90.2\% of the domains with \texttt{A} records have an active website (5.1\,M out of 5.6\,M, in Table \ref{tab:datasets}) and 91.1\% with \texttt{AAAA} records have an active IPv6 website~(2.58\,M out of 2.83\,M, in Table \ref{tab:datasets-ipv6}). Using Internet-wide port-scan data from Rapid7 Sonar, we analyzed the number of domains that have both an IPv4 address and an open HTTP(s) port. The results from this preliminary analysis show that this applies to 90.7\% of domains considered, on average per TLD.
        
\textbf{CNAMEs}. For cases in which a domain name uses a \texttt{CNAME} rather than \texttt{A} or \texttt{AAAA} records  -- as typically used with domains hosted in content-delivery networks (CDNs) -- we fully expand the \texttt{CNAME} chain until it is terminated by \texttt{A} or \texttt{AAAA} records. We then map the domain name in question to the AS number(s) at the end of the chain.
        
\textbf{Parked domains}. In case a domain name leads to non-responsive, parking, or redirecting website, we still take into account the infrastructure where the website is supposedly hosted -- classifying this typology of websites is left as future work.
    
\textbf{Subdomains}. Given that OpenINTEL primarily knows zone files and the second-level domain names therein, we mostly infer hosting at the domain apex and for the third-level \dns{www.} label (except for \texttt{CNAME} records, where arbitrary labels can be encountered and are expanded). 

\textbf{Single vantage-point}. OpenINTEL collects measurements from a single vantage point in the Netherlands, which can introduce a bias towards ``nearby'' \texttt{A} and \texttt{AAAA} records when DNS-based load balancing is used. This bias can, in theory, increase the concentration towards European IP address space. However, even if hosting companies operate address space in several countries, the AS number of their headquarter country remains the same.

\subsection{Data set}
To measure the evolution of web hosting, we use historical data from OpenINTEL -- one day per year (on May 5th), covering the last five years (2017 -- 2021). Table \ref{tab:all-datasets} summarizes the resulting data set, for all TLDs under consideration combined. We measure 11 -- 19 TLDs, covering 161\,M -- 201\,M domains, hosted by 35\,k -- 40\,k IPv4 ASes and fewer than 5\,k IPv6 ASes. 


\begin{table}
    \centering
    \resizebox{\linewidth}{!}{
    	\setlength{\tabcolsep}{1.5pt}
    	\begin{tabular}{c|c|c|c|r|c|c}
            \small\textbf{Year} & \textbf{TLDs} & \textbf{Domains} & \textbf{IPv4} & \textbf{IPv6} & \textbf{ASes(v4)} & \textbf{ASes(v6)} \\ \hline
            2017 & 11 & 161.4\,M & 149.1\,M &  8.7\,M & 35.7\,k & 3.4\,k \\ \hline
            2018 & 13 & 175.0\,M & 157.9\,M & 11.9\,M & 38.2\,k & 3.8\,k \\ \hline
            2019 & 14 & 180.6\,M & 166.3\,M & 15.0\,M & 38.8\,k & 4.0\,k \\ \hline
            2020 & 17 & 188.5\,M & 173.0\,M & 17.7\,M & 39.6\,k & 4.3\,k \\ \hline
            2021 & 19 & 200.7\,M & 182.6\,M & 21.8\,M & 40.4\,k & 4.6\,k
        \end{tabular}%
    }
	\caption{Aggregated data sets, measured each year on May 5.}
	\label{tab:all-datasets}
\end{table}

%% file: src/hosting_concentration.tex
\section{Web hosting concentration}
\label{sec:web-conc}

In this section, we focus on determining the largest hosting ASes in our data sets. Table~\ref{tab:all-datasets} summarized the aggregate data. Our goal is to determine if there is a significant concentration in the domain namespace. To do that, we compute the percentage of the Top 5, 10, 20, and 100 hosting ASes per year.

\subsection{Concentration on Top hosting providers}

\begin{figure}[ht!]
    \centering
    \begin{subfigure}[t]{0.5\textwidth}
        \centering
        \includegraphics[width=\linewidth]{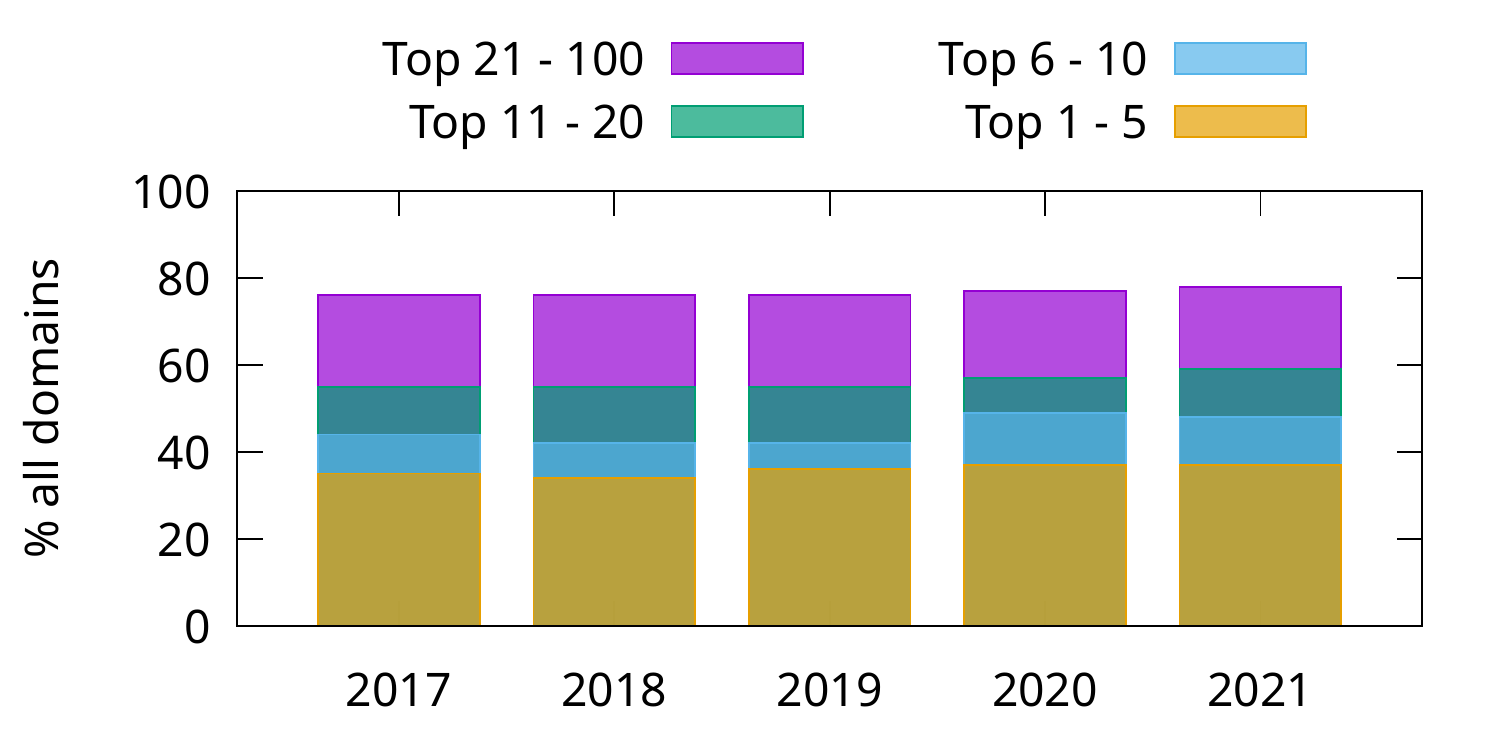}
        \caption{IPv4}
        \label{fig:all-top10-ipv4}
    \end{subfigure}
    \begin{subfigure}[t]{0.5\textwidth}
        \centering
        \includegraphics[width=\linewidth]{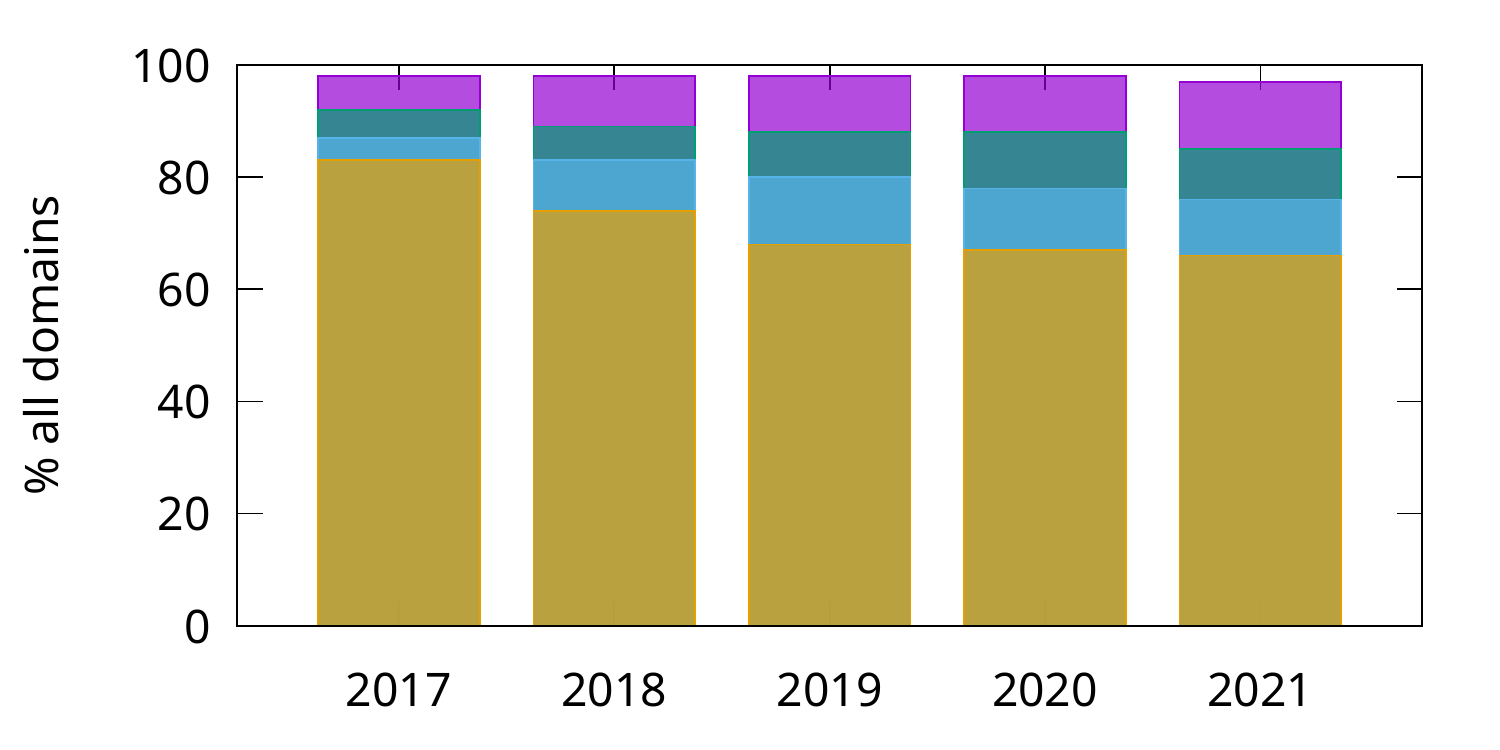}
        \caption{IPv6}
        \label{fig:all-top10-ipv6}
    \end{subfigure}
    \caption{Concentration in Top hosting ASes}
    \label{fig:all-top10}
\end{figure}

Figure \ref{fig:all-top10} shows the result of our analysis. The x-axis contains the considered year (2017 through 2021) and the y-axis shows the percentage of domains hosted in Top hosting ASes. For IPv4 (Fig.~\ref{fig:all-top10-ipv4}), the Top 5 hosting ASes host more than one-third of the 120\,M -- 150\,M domains (depending on the year) that had IPv4 records. We observe that this concentration has been slightly growing over the last five years. Note that our data contains 35\,k -- 40\,k hosting ASes for the evaluated years~(Table~\ref{tab:all-datasets}), which shows how significant the concentration in top hosting ASes is. For IPv6~(Figure \ref{fig:all-top10-ipv6}), the Top 5 hosting ASes account for 64--71\% of the 10\,M -- 21\,M domain names with IPv6 records. This reinforces the notion that big tech companies have committed to fully implement IPv6 to encourage smaller companies to make the transition to IPv6 \cite{worldipv6} and that, despite the still low adoption of IPv6, these companies remain at the forefront until today.

In general we observe that, for IPv4, the concentration has been relatively stable since 2017, and the percentage of the domains hosted by the Top 10 ASes has slightly increased. However, for IPv6, despite being far more concentrated, there appears to be a reverse trend, most likely related to the adoption of IPv6 by more hosting providers (3.4\,k to 4.6\,k ASes between 2017 and 2021--Table~\ref{tab:all-datasets}).

\subsection{Ranking the Top hosting providers}
\label{sec:google}

Along the years, the hosting market can present movement of domains, migrating from one hosting provider to another. In order to observe how the market has been dominated by the Top providers, we take into account both the total number of domains hosted by each provider per year, as well as the percentage of the market occupied by each provider.

\begin{figure}[h!]
    \centering
    \begin{subfigure}[t]{0.5\textwidth}
        \centering
        \includegraphics[width=\linewidth]{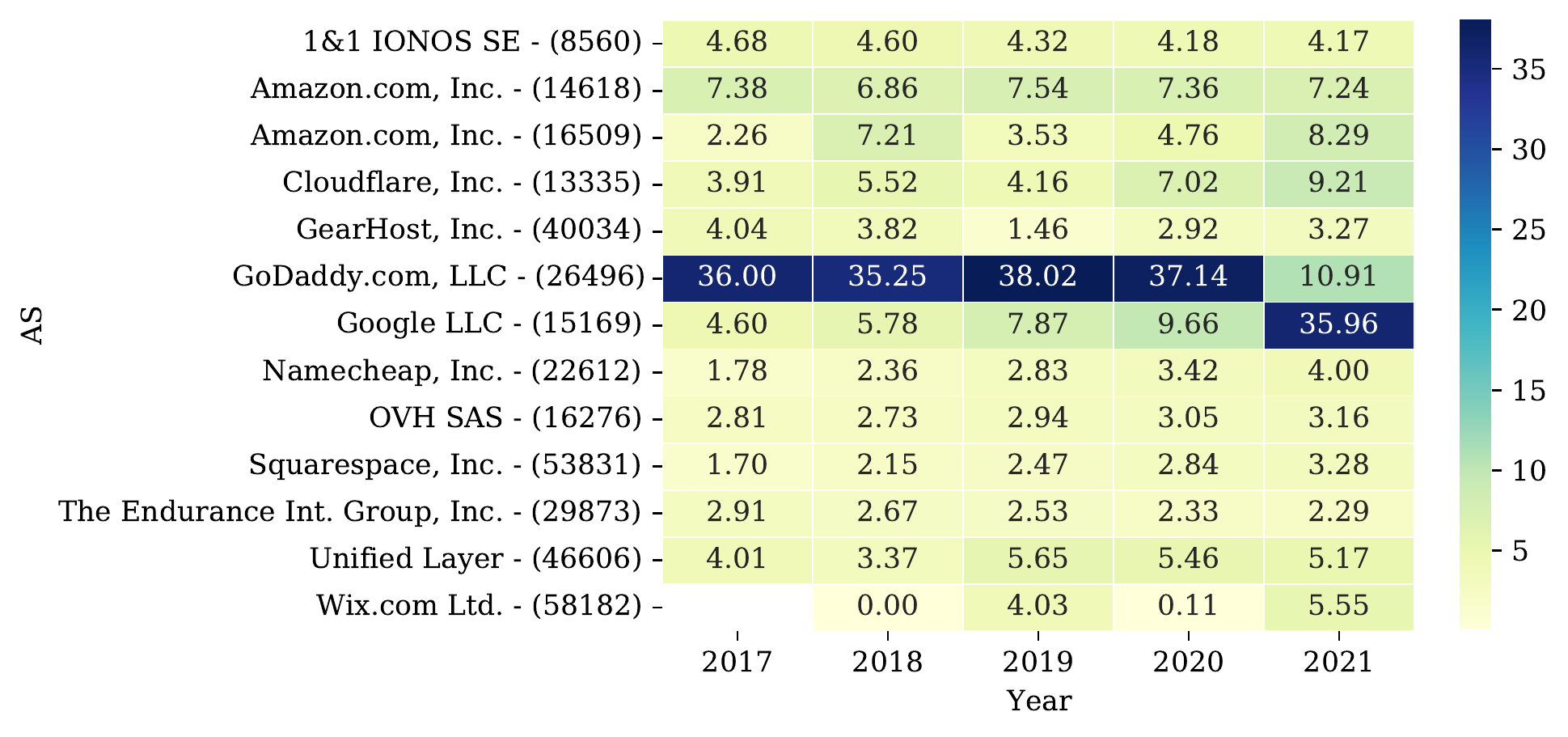}
        \caption{IPv4}
        \label{fig:v4-top10ASes}
    \end{subfigure}
    \begin{subfigure}[t]{0.5\textwidth}
        \centering
        \includegraphics[width=\linewidth]{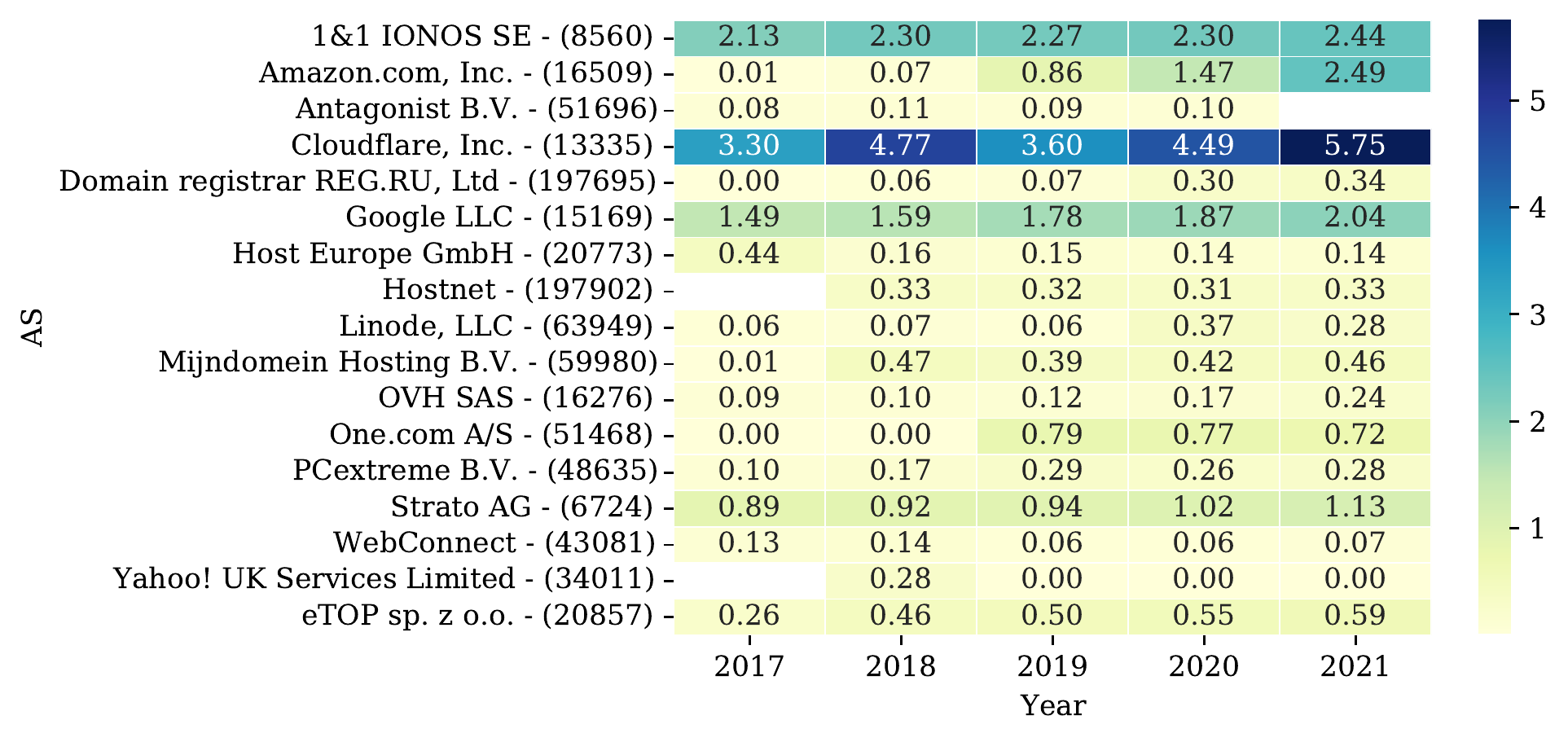}
        \caption{IPv6}
        \label{fig:v6-top10ASes}
    \end{subfigure}
    \caption{Top Hosting ASes evolution}
   \label{fig:top10ASes}
\end{figure}

\begin{figure}[h!]
    \centering
    \begin{subfigure}[t]{0.5\textwidth}
        \centering
        \includegraphics[width=\linewidth]{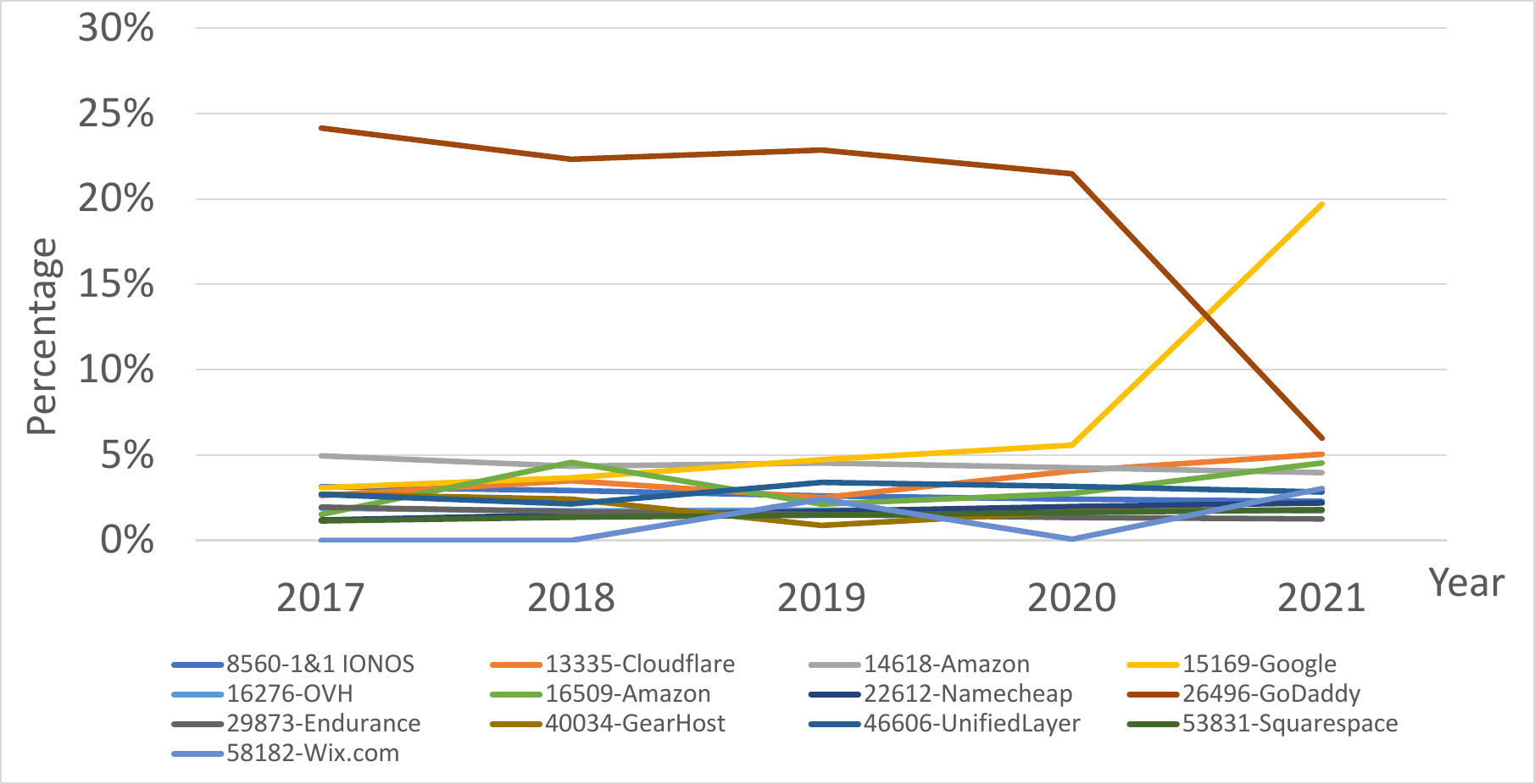}
        \caption{IPv4}
        \label{fig:TopHostingPerYear-ipv4}
    \end{subfigure}
    \begin{subfigure}[t]{0.5\textwidth}
        \centering
        \includegraphics[width=\linewidth]{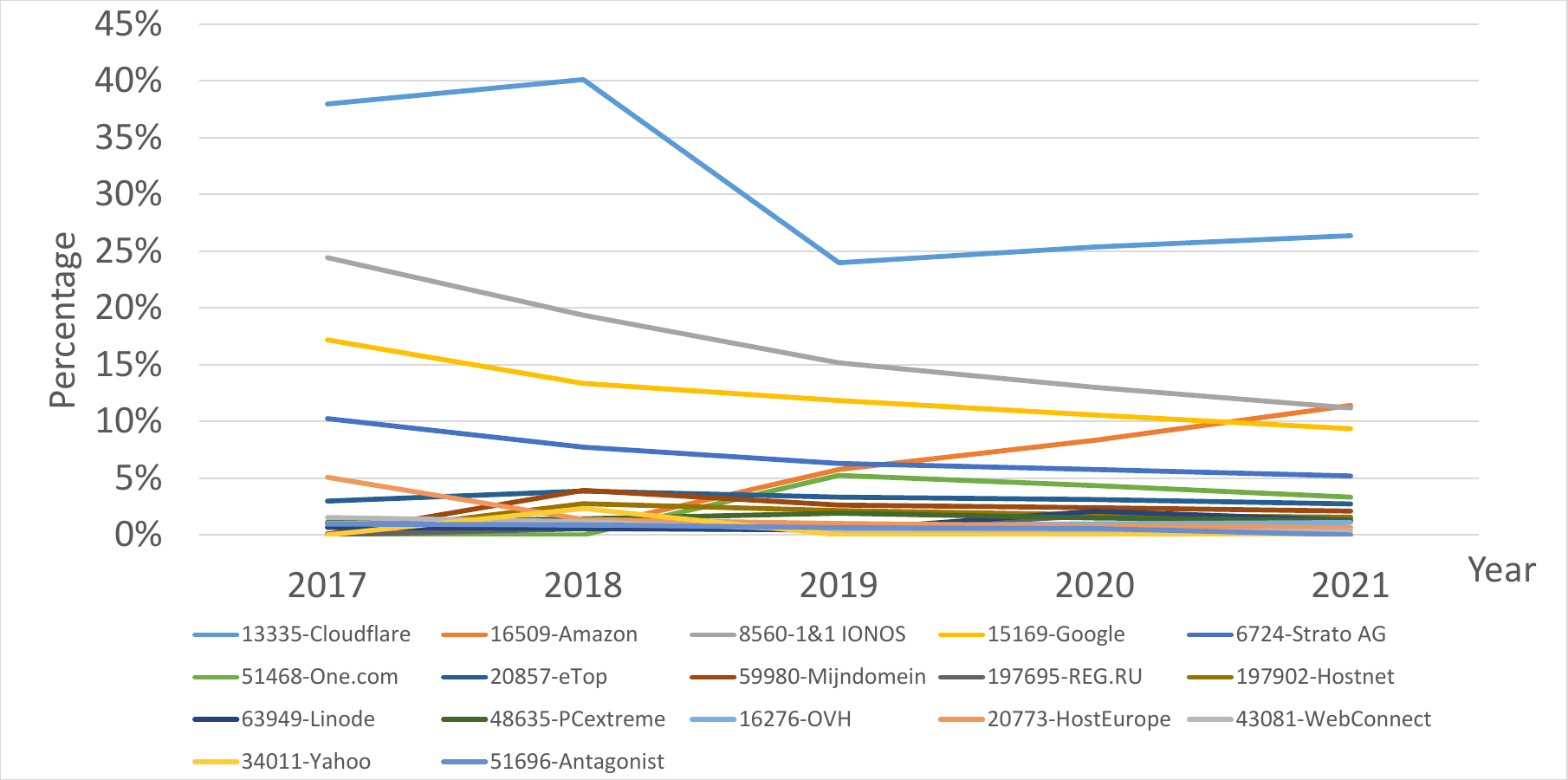}
        \caption{IPv6}
        \label{fig:fig:TopHostingPerYear-ipv6}
    \end{subfigure}
    \caption{Percentage of domains hosted by Top Hosting ASes per Year}
   \label{fig:TopHostingPerYear}
\end{figure}

Google's dominance of the hosting market occurred only in 2021. Figures \ref{fig:top10ASes} and \ref{fig:TopHostingPerYear} show the evolution of the Top hosting ASes observed from 2017 to 2021, in terms of \textit{millions of hosted domain names} and \textit{percentage of hosted domains}, respectively. As can be observed, most ASes grow in terms of hosted domains and ratio. The major surprise is the swap of positions of Google (AS15169) and Cloudflare (AS26496). In 2020, Cloudflare hosted 37.14 million domain names, but only 10.91\,M in 2021. In the same period, Google showed growth: from 9.66\,M to 35.96\,M domains.


To investigate the reasons for this inversion, we compared both 2020 and 2021 results and found that 
17\,M domains hosted by Cloudflare in 2020 migrated to Google Cloud. We investigated these domains and found that they are all hosted on the same IPv4 address. This IPv4 address hosts a simple template page for a parked domain; therefore, we marked all the domains as \textit{parked} (parked domains are a part of the domain industry in which domains are parked to reserve, resell, or run ads for profit~\cite{vissers2015parking}). 

This case exemplifies the amount of power that lies in the hands of a large hosting provider as it can shift the entire domain distribution with a single policy decision. 

For IPV6 (Figure \ref{fig:fig:TopHostingPerYear-ipv6}), this behavior is not repeated. Cloudflare, which is the provider with the highest number of domains, went from 40\% to 26\% of domains. We note that despite the providers having shown an increase in the number of hosted domains (figure \ref{fig:v6-top10ASes}), they showed a decrease in the percentage of concentration in relation to the total number of domains (figure \ref{fig:fig:TopHostingPerYear-ipv6}) from 2017 to 2021.

%% file: src/concentration_per_TLD.tex
\section{Concentration per TLD}
\label{sec:results}

Now we move to evaluate the domain concentration per TLD to determine if distinct TLDs have similar concentration patterns. We start by splitting our data set from Table~\ref{tab:all-datasets} into TLDs and next evaluate each. Then we compute the concentration of the Top 5, 10, and 20 hosting ASes, similar to Section \ref{sec:web-conc}.

Table~\ref{tab:datasets-ipv4} shows the data set split per TLD, for the 2021 IPv4 measurements. Table~\ref{tab:datasets-ipv6} shows IPv6. The tables provide a detailed overview of characteristics of the domains present in each of the TLDs. The row \texttt{Domains} shows the number of domains in the zone file. The \texttt{Responsive} row represents the domains that respond to \textit{SOA} query (that means well configured). We see that for all zones, the majority of domains have \texttt{A} records associated with them (\texttt{with hosting} column), either on the \dns{apex} or under \dns{www}. Interestingly, some zones have more \texttt{A} records on the \textit{apex} and others more on \dns{www}. For example, for \dns{.at}, 99\% of domains with hosting have \texttt{A} record under \dns{www}, but only 96.8\% on the \dns{apex}. For \dns{.ru}, these values are 99.7\% and 96.9\%, respectively. The \dns{apex} row is the subset of \texttt{with hosting} that have hosting on the \dns{apex}. Row \dns{www} represents the subset that has hosting on \dns{www}. The row with the symbol "$\cap$" represents the domains that have the same \texttt{A} records for both \dns{apex} and \dns{www}. The symbol $!=$ means that the domains have different \texttt{A} records for the \dns{apex} and \dns{www}. Finally, the last three rows represent the domains that are under a single, multiple, or unique ASes, respectively.

The \texttt{A} records defined at the \dns{apex} and \dns{www} labels may \textit{differ}. For example, the domain \dns{klm.de} has \texttt{171.21.122.79} (AS28806) as an \texttt{A} record on the apex. At the same time, \dns{www.klm.de} has a \texttt{CNAME}~\cite{rfc1034} that points to \texttt{23.45.75.2}, which is a multi-homed address~\cite{rfc4116}, announced by two ASes: AS20940 and AS16625, both belonging to Akamai, a large CDN provider. On a browser, using either name leads the user to the same website. As we show in Table \ref{tab:datasets}, the majority of domains have the same \texttt{A} records in both \dns{apex} and \dns{www} (row "\texttt{$\cap$}"), but from 3.7\% to 10.7\% have different (row $!=$), depending on the ccTLD. Each domain name may use multiple ASes to announce its \texttt{A} records to achieve load-balancing and resilience against failure. However, fewer than 5\% of all domains use multiple ASes for hosting for the evaluated TLDs (Table~\ref{tab:datasets}).

\begin{table*}[!h]
  \centering
  \begin{subtable}[t]{1\linewidth}
   \centering
    \resizebox{\linewidth}{!}{
	    \setlength{\tabcolsep}{.7pt}
        \begin{tabular}{c|r|r|r|r|r|r|r|r|r|r|r|r|r|r|r|r|r|r|r}
         & \multicolumn{16}{c|}{ccTLDs} & \multicolumn{3}{c|}{gTLDs} \\
         & \textbf{.at} & \textbf{.ca} & \textbf{.ch} & \textbf{.co} & \textbf{.dk} & \textbf{.ee} & \textbf{.fi} & \textbf{.gt} & \textbf{.li} & \textbf{.na} & \textbf{.nl} & \textbf{.nu} & \textbf{.ru} & $\mathbf{.p\phi}$ & \textbf{.se} & \textbf{.us} & \textbf{.com} & \textbf{.net} & \textbf{.org} \\ \hline
        Domains & 1.36\,M & 3.03\,M & 2.25\,M & 2.86\,M & 1.34\,M & 139.20\,k & 510.94\,k & 19.90\,k & 60.91\,k & 5.21\,k & 6.17\,M & 252.21\,k & 4.73\,M & 630.71\,k & 1.49\,M & 1.65\,M & 150.88\,M & 13.05\,M & 10.17\,M \\ \hline
        Responsive & 1.31\,M & 2.84\,M & 2.07\,M & 2.68\,M & 1.22\,M & 135.52\,k & 476.80\,k & 17.39\,k & 55.11\,k & 4.84\,k & 5.84\,M & 233.21\,k & 4.47\,M & 579.72\,k & 1.41\,M & 1.49\,M & 142.21\,M & 11.99\,M & 9.58\,M \\ \hline
        with hosting & 1.24\,M & 2.78\,M & 1.93\,M & 2.62\,M & 1.11\,M & 130.56\,k & 442.48\,k & 16.03\,k & 50.54\,k & 3.88\,k & 5.67\,M & 221.43\,k & 4.33\,M & 557.21\,k & 1.34\,M & 1.49\,M & 138.11\,M & 11.27\,M & 9.26\,M \\ \hline
        apex & 1.20\,M & 2.76\,M & 1.91\,M & 2.60\,M & 1.09\,M & 129.21\,k & 431.50\,k & 15.71\,k & 49.06\,k & 3.43\,k & 5.59\,M & 217.47\,k & 4.32\,M & 556.08\,k & 1.32\,M & 1.47\,M & 136.38\,M & 11.07\,M & 9.16\,M \\ \hline
        \dns{www.} & 1.23\,M & 2.73\,M & 1.91\,M & 2.56\,M & 1.08\,M & 128.59\,k & 435.52\,k & 15.59\,k & 48.75\,k & 3.82\,k & 5.62\,M & 219\,k & 4.20\,M & 536.30\,k & 1.33\,M & 1.45\,M & 135.87\,M & 11.03\,M & 9.13\,M \\ \hline
        $\cap$ & 1.11\,M & 2.46\,M & 1.69\,M & 2.26\,M & 1.01\,M & 12.21\,k & 386.72\,k & 13.92\,k & 43.74\,k & 3.10\,k & 5.34\,M & 206.88\,k & 3.81\,M & 511.60\,k & 1.23\,M & 1.32\,M & 118.31\,M & 9.79\,M & 8.06\,M \\ \hline
        $!=$ & 77.05\,k & 258.14\,k & 187.12\,k & 282.24\,k & 53.57\,k & 5.10\,k & 37.83\,k & 1.35\,k & 3.53\,k & 267 & 213.69\,k & 8.16\,k & 378.45\,k & 23.57\,k & 76.17\,k & 121.47\,k & 15.82\,M & 1.03\,M & 970.85\,k \\ \hline
        Single & 1.21\,M & 2.68\,M & 1.85\,M & 2.53\,M & 1.08\,M & 127.61\,k & 427.84\,k & 15.27\,k & 48.97\,k & 3.73\,k & 5.58\,M & 216.28\,k & 4.26\,M & 547.62\,k & 1.3\,M & 1.44\,M & 131.37\,M & 10.81\,M & 8.86\,M \\ \hline
        Multiple & 32.97\,k & 103.53\,k & 74.53\,k & 91.33\,k & 25.04\,k & 2.95\,k & 14.63\,k & 761 & 1.56\,k & 153 & 90.35\,k & 5.14\,k & 72.78\,k & 9.58\,k & 42.82\,k & 45.75\,k & 6.74\,M & 454.53\,k & 395.45\,k \\ \hline
        Unique ASes & 4.22\,k & 4.74\,k & 4.59\,k & 6.70\,k & 3.05\,k & 1.53\,k & 2.35\,k & 631 & 1.80\,k & 255 & 4.77\,k & 1.89\,k & 7.93\,k & 2.72\,k & 3.69\,k & 7.64\,k & 35.15\,k & 23.41\,k & 19.51\,k
        \end{tabular}%
    }
	\caption{IPv4}
	\label{tab:datasets-ipv4}
    \bigskip
  \end{subtable}
  \begin{subtable}[t]{1\linewidth}
    \centering
    \resizebox{\linewidth}{!}{
	    \setlength{\tabcolsep}{.7pt}
        \begin{tabular}{c|r|r|r|r|r|r|r|r|r|r|r|r|r|r|r|r|r|r|r}
            & \multicolumn{16}{c|}{ccTLDs} & \multicolumn{3}{c|}{gTLDs} \\
             & \textbf{.at} & \textbf{.ca} & \textbf{.ch} & \textbf{.co} & \textbf{.dk} & \textbf{.ee} & \textbf{.fi} & \textbf{.gt} & \textbf{.li} & \textbf{.na} & \textbf{.nl} & \textbf{.nu} & \textbf{.ru} & \textbf{.se} & \textbf{.us} & $\mathbf{.p\phi}$ & \textbf{.com} & \textbf{.net} & \textbf{.org} \\ \hline
        Domains      & 1.36M        & 3.03M        & 2.25M        & 2.86M        & 1.34M        & 139k         & 510k         & 19.90k       & 60.91k       & 5.21k        & 6.17M        & 252k         & 4.73M        & 630k         & 1.49M        & 1.65M             & 150.88M       & 13.05M        & 10.17M        \\ \hline
        Responsive & 1.31\,M & 2.84\,M & 2.07\,M & 2.68\,M & 1.22\,M & 135.52\,k & 476.80\,k & 17.39\,k & 55.11\,k & 4.84\,k & 5.84\,M & 233.21\,k & 4.47\,M & 579.72\,k & 1.41\,M & 1.49\,M & 142.21\,M & 11.99\,M & 9.58\,M \\ \hline
        with hosting & 197.36k      & 150.47k      & 515.70k      & 336.12k      & 268.08k      & 7.97k        & 54.25k       & 1.63k        & 10.80k       & 247          & 2.83M        & 62.37k       & 895.99k      & 101.73k      & 366.84k      & 107.33k           & 13.72M        & 1.12M         & 1M            \\ \hline
        apex         & 190.95k      & 128.13k      & 491.77k      & 309.57k      & 258.74k      & 7.21k        & 48.12k       & 1.39k        & 10.29k       & 206          & 2.76M        & 60.78k       & 887.69k      & 101.20k      & 355.05k      & 96.71k            & 12.03M        & 995.07k       & 874.55k       \\ \hline
        www.         & 191.53k      & 145.81k      & 501.81k      & 321.67k      & 255.91k      & 7.68k        & 52.63k       & 1.54k        & 10.28k       & 237          & 2.77M        & 61.25k       & 857.07k      & 100.01k      & 362.03k      & 95.75k            & 13.19M        & 1.06M         & 968k          \\ \hline
        $\cap$       & 183.59k      & 108.94k      & 474.58k      & 284.07k      & 245.32k      & 6.81k        & 45.63k       & 1.25k        & 9.64k        & 190          & 2.69M        & 59.46k       & 847k         & 99.39k       & 348.89k      & 73.99k            & 10.75M        & 879.61k       & 767.34k       \\ \hline
        $!=$         & 1.53k        & 14.54k       & 3.30k        & 11.04k       & 1.25k        & 112          & 862          & 37           & 131          & 6            & 8.17k        & 198          & 1.20k        & 95           & 1.35k        & 11.13k            & 753.03k       & 58.06k        & 72.56k        \\ \hline
        Single       & 196.66k      & 148.54k      & 513.89k      & 333.53k      & 267.03k      & 7.89k        & 53.43k       & 1.62k        & 10.72k       & 245          & 2.82M        & 62.14k       & 895.01k      & 101.58k      & 365.75k      & 105.94k           & 13.59M        & 1.11M         & 995.60k       \\ \cline{1-18} \cline{20-20} 
        Multiple     & 696          & 1.92k        & 1.81k        & 2.59k        & 1.05k        & 81           & 821          & 10           & 81           & 2            & 9.98k        & 236          & 986          & 149          & 1.08k        & 1.38k             & 131.11k       & 7.90k         & 7.87k         \\ \hline
        Unique ASes  & 591          & 405          & 678          & 604          & 366          & 178          & 314          & 53           & 266          & 23           & 669          & 330          & 651          & 174          & 522          & 594               & 3.4k          & 2.73k         & 2.03k        
        \end{tabular}
    }
	\caption{IPv6}
	\label{tab:datasets-ipv6}
  \end{subtable}
  \caption{Data sets 2021-05-05}
  \label{tab:datasets}
\end{table*}

We study hosting concentration per TLD to investigate if the concentration is uniform or dissimilar. We use per-TLD data for 2021 and then compute the concentration of the top 5, 10, and 20 hosting ASes.

\begin{figure}[h!]
    \centering
    \begin{subfigure}[t]{0.5\textwidth}
        \centering
        \includegraphics[width=0.8\linewidth]{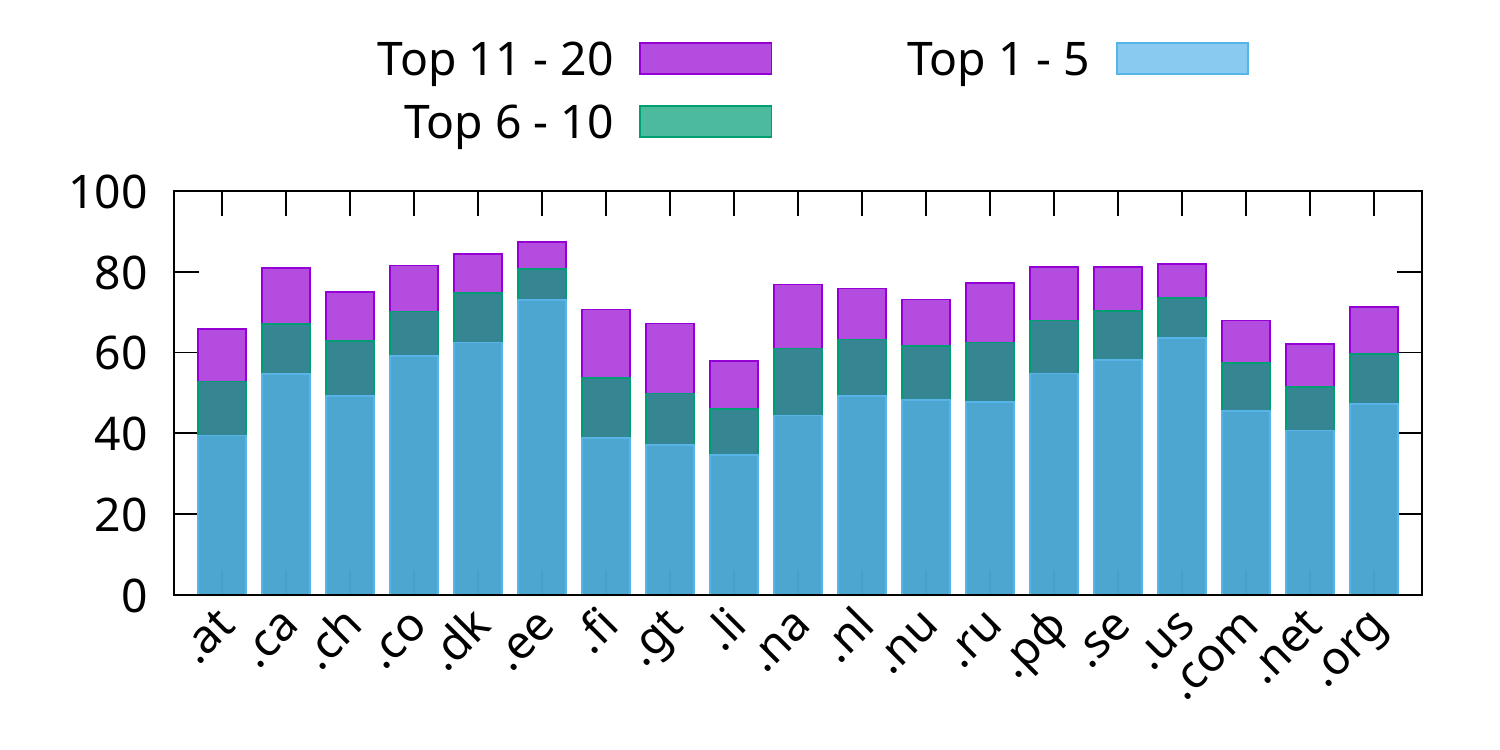}
        \caption{IPv4}
        \label{fig:conc-top10-ipv4}
    \end{subfigure}
    \begin{subfigure}[t]{0.5\textwidth}
        \centering
        \includegraphics[width=0.8\linewidth]{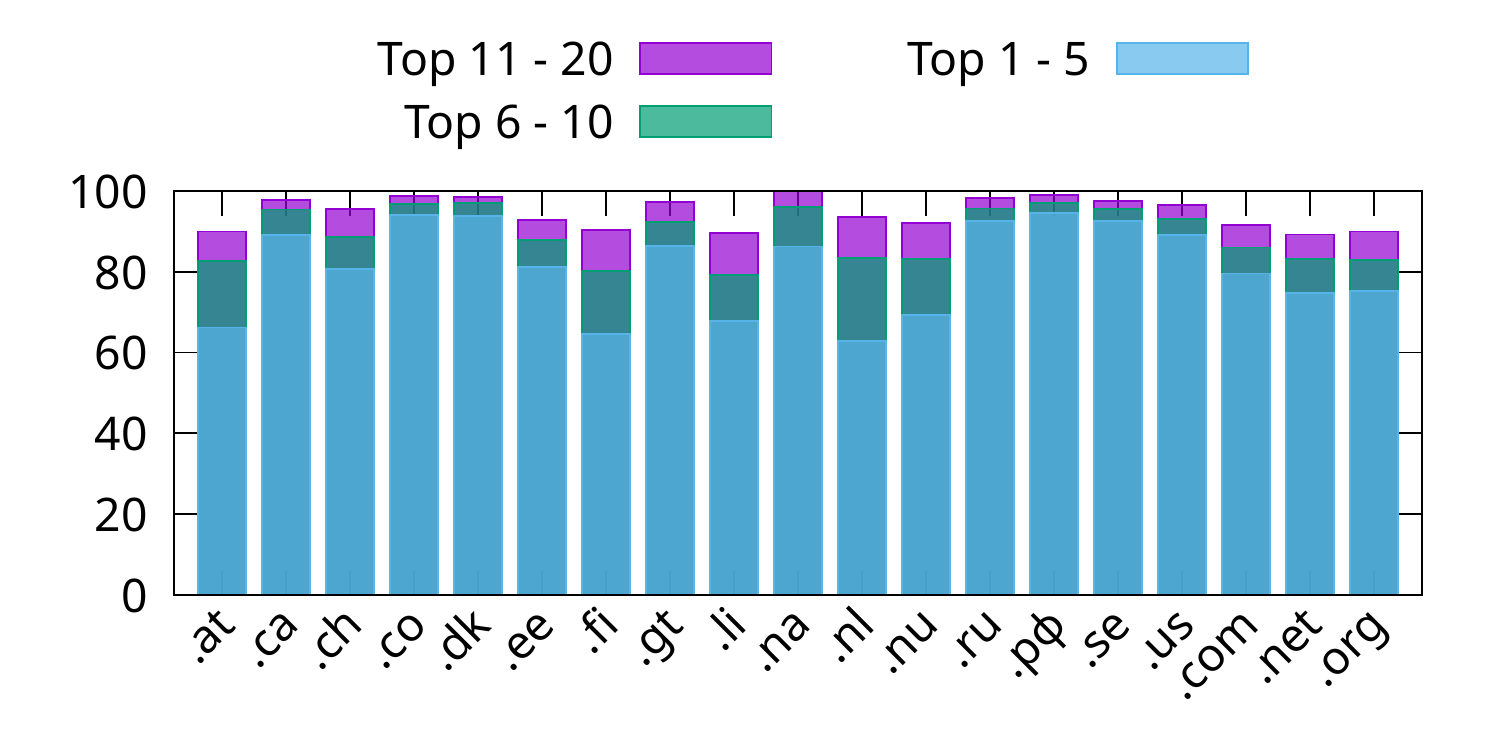}
        \caption{IPv6}
        \label{fig:conc-top10-ipv6}
    \end{subfigure}
    \caption{Hosting concentration 2021 per TLD}
   \label{fig:conc-top10}
\end{figure} 


\subsection{Web hosting concentration}
\label{sec:hosting-concentration}

Figure~\ref{fig:conc-top10} for each TLD and its respective top hosting ASes the percentage of domain names. A significant concentration is apparent for all TLDs:  the Top 5 hosting ASes are responsible for 37--73\% of the domains, depending on the TLD, for IPv4 (Figure \ref{fig:conc-top10-ipv4}). Estonia's \dns{.ee} has the highest concentration level: 73\%. For IPv6, the concentration is even more significant: from 62--94\%. We also observe that the difference between the top 20 hosting ASes and the Top 5 is not that significant, which further demonstrates how concentrated the market is. We also see no significant differences between ccTLD and gTLD concentration.

\begin{figure}[h!]
    \centering
    \begin{subfigure}[t]{0.5\textwidth}
        \centering
        \includegraphics[width=0.8\linewidth]{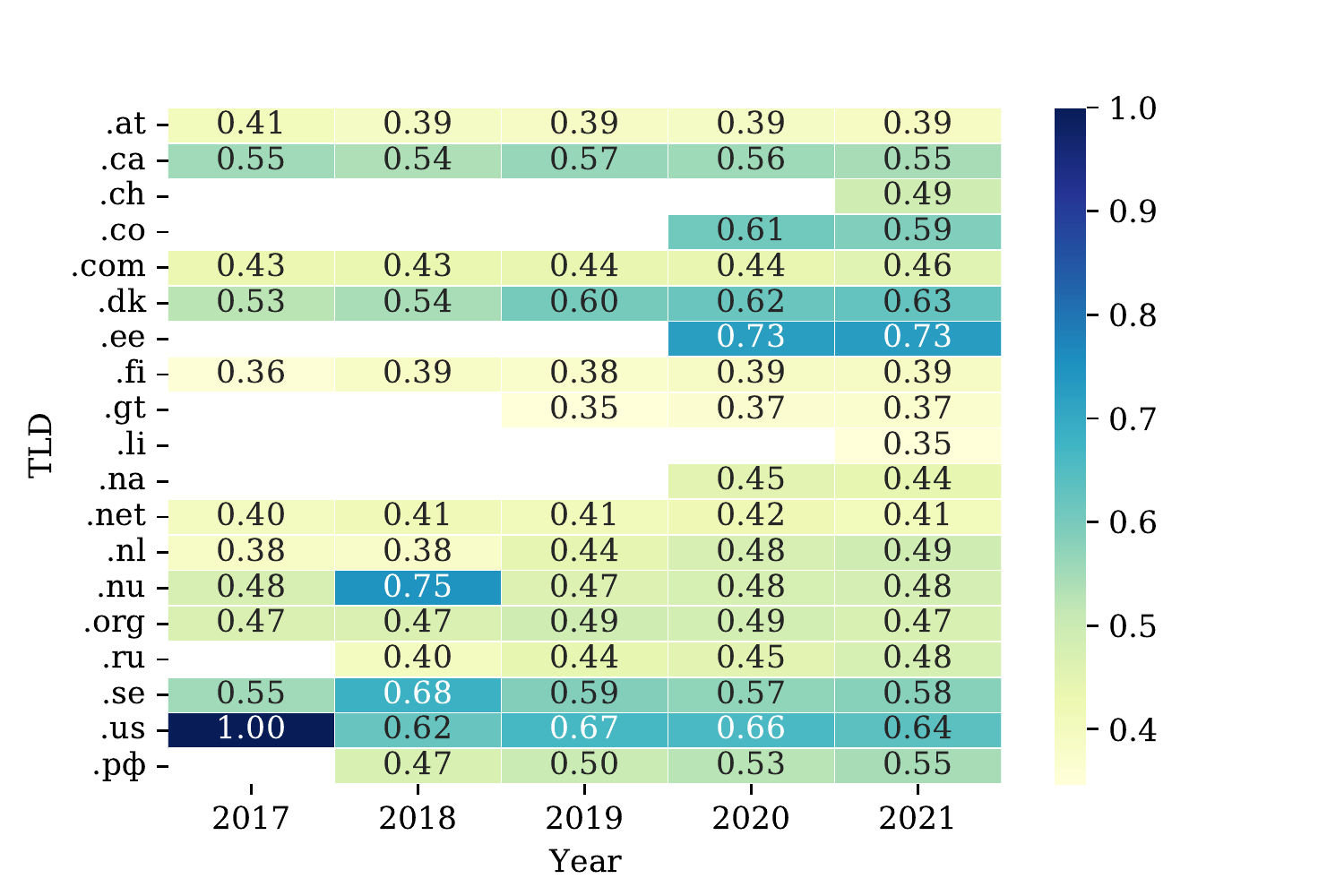}
        \caption{IPv4}
        \label{fig:v4-top5-time}
    \end{subfigure}
    \begin{subfigure}[t]{0.5\textwidth}
        \centering
        \includegraphics[width=0.8\linewidth]{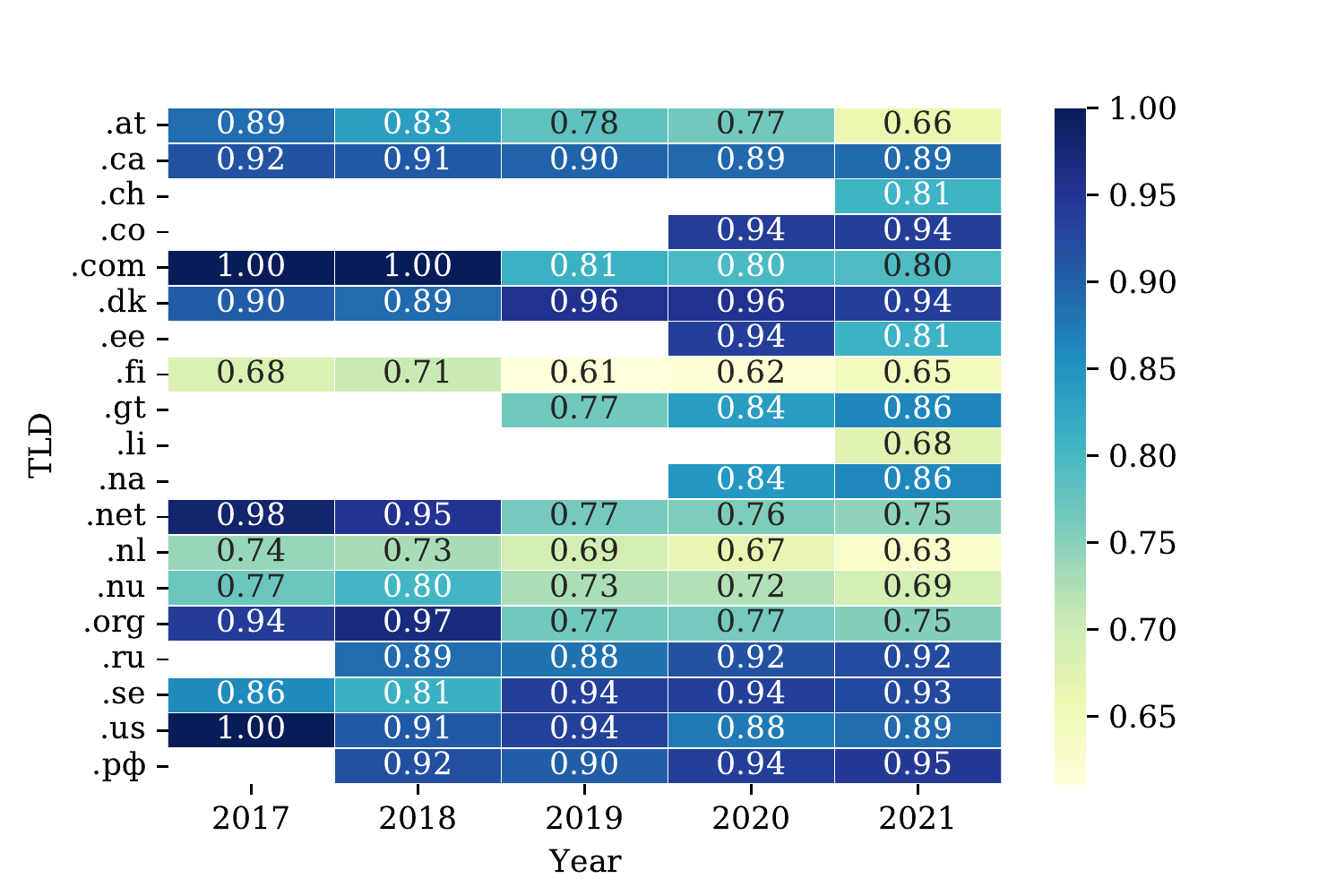}
        \caption{IPv6}
        \label{fig:v6-top5-time}
    \end{subfigure}
    \caption{Ratio of domains hosted by Top 5 ASes per TLD and Year}
   \label{fig:top5-time}
\end{figure}

\textit{Changes over time:} Figure \ref{fig:v4-top5-time} shows the variations in concentration of hosting for the Top 5 hosting ASes, per TLD in terms of the percentage of SLDs hosted.  We observe that this concentration has been rather stable for most TLDs for IPv4 (Figure \ref{fig:v4-top5-time}). This is different from the IPv6 case (Figure \ref{fig:v6-top5-time}), which appears to have reduced the concentration of domains in the top 5 providers in some TLDs. For IPv4 some ccTLDs have experienced an increasing concentration: Denmark's \dns{.dk} has gone from 53\% to 63\% of the domains being hosted by five providers in the last five years. \dns{.nl}, similarly, went from 38\% to 49\%. For IPv6, we notice that part of the TLDs reduce the ratio of domains hosted by Top 5 ASes. As we can see, Austria's \dns{.at} went from 89\% to 66\%. As well, \dns{.net} went from 98\% to 75\%.

\subsection{Hosting Providers Location}
\label{sec:hosting-location}

Next, we investigate the locations of hosting companies. For each AS used for hosting, we lookup the company's origin and then map the domain names they host to this country. Note from earlier discussion that this is not the same as geolocating the hosting IP addresses (section~\ref{sec:methodology}).

For each TLD we count the number of domains per country and classify each country into four categories, in the following priority order: \textit{local} (if the country is the same as the TLD), US-based (if the the country is the US, given most of large cloud companies are from the US), same official language (for example, \dns{.ca} and \dns{.mt} also use English as one their official languages), and rest (the remaining domains). Note that each AS/country gets only a label, even if multiple may apply. For example, AS15169 is both local and US hosting for \dns{.us}, but we consider it only as local for \dns{.us}, given that local takes priority in our classification.

\begin{figure}[h!]
    \centering
    \begin{subfigure}[t]{0.5\textwidth}
        \centering
        \includegraphics[width=0.8\linewidth]{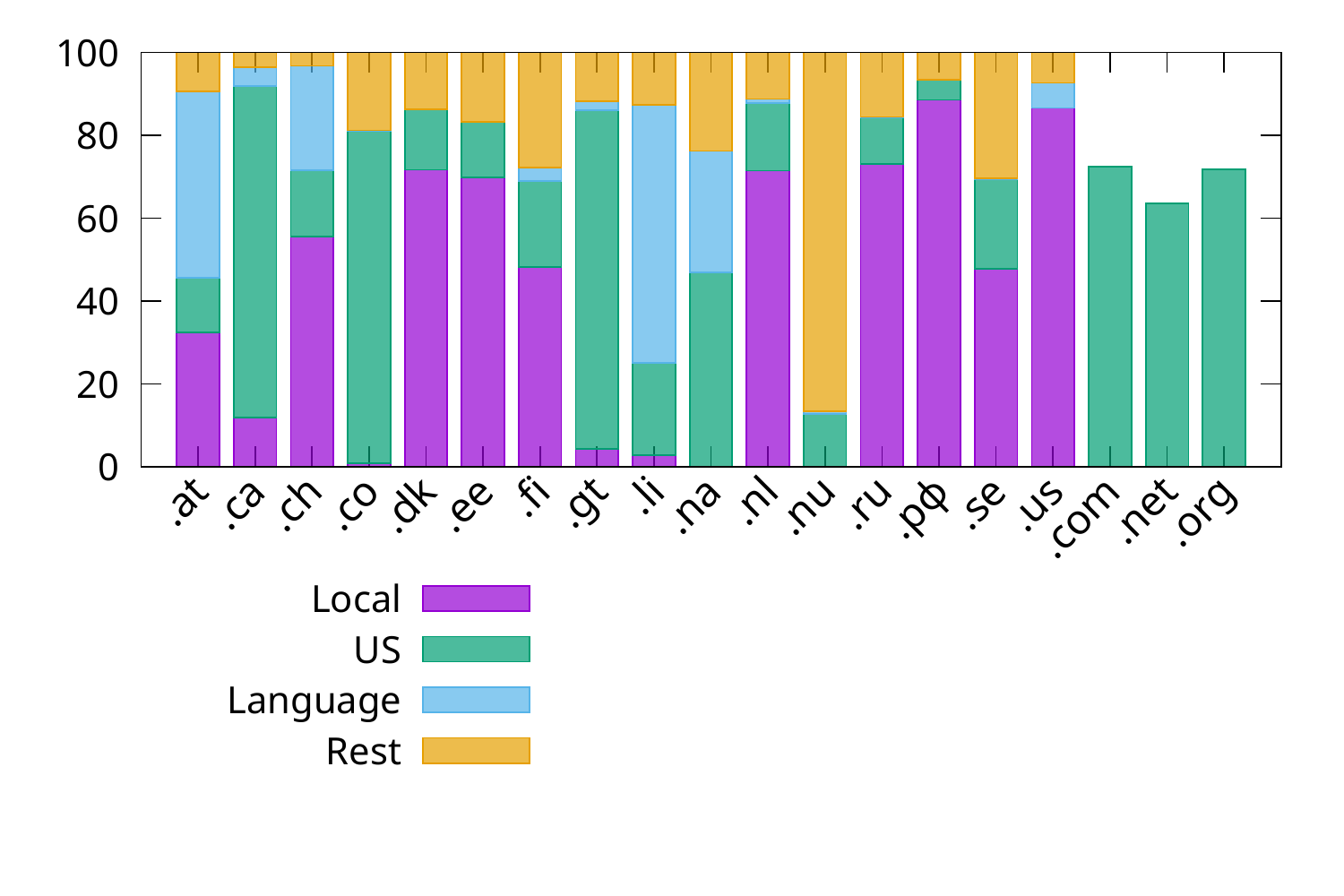}
        \caption{IPv4}
        \label{fig:conc-type-ipv4}
    \end{subfigure}
    \begin{subfigure}[t]{0.5\textwidth}
        \centering
        \includegraphics[width=0.8\linewidth]{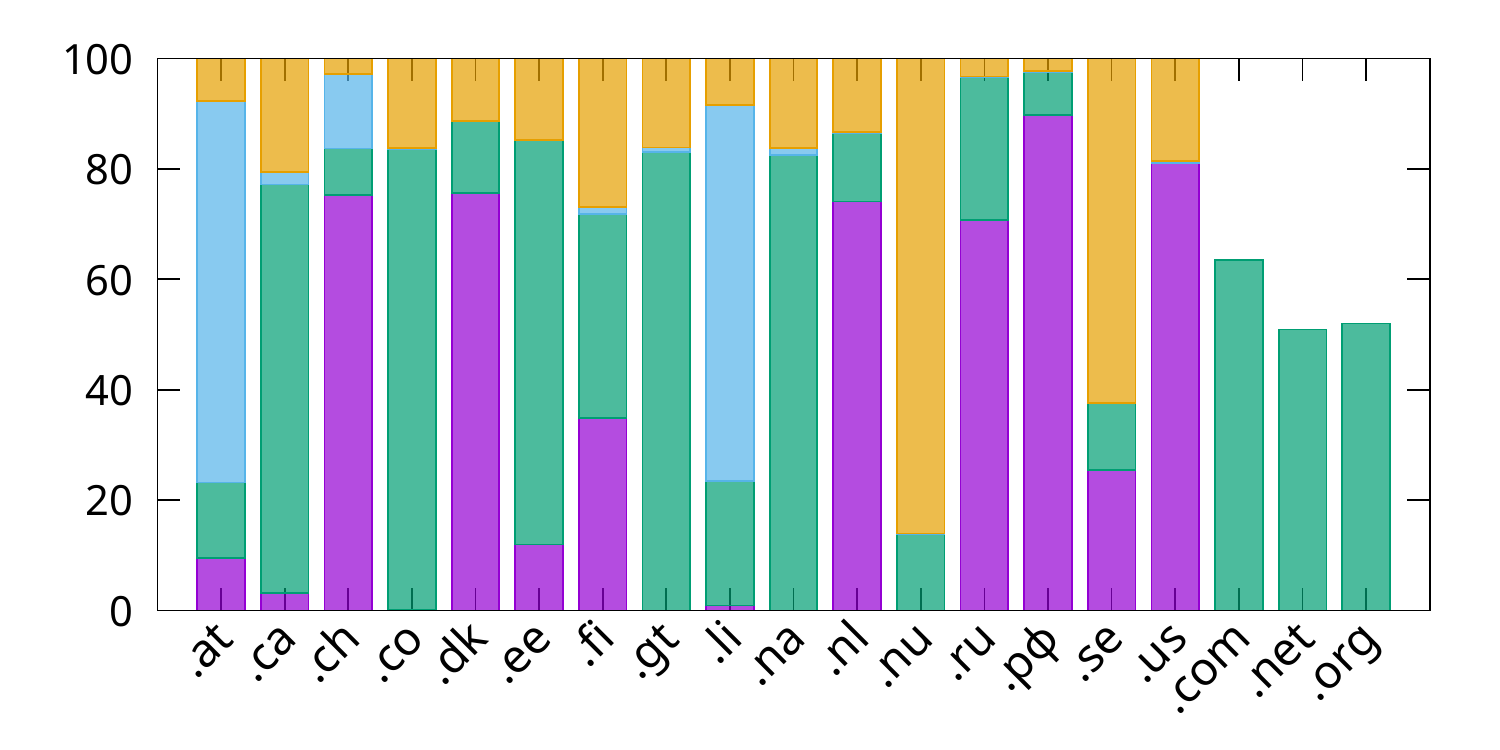}
        \caption{IPv6}
        \label{fig:conc-type-ipv6}
    \end{subfigure}
    \caption{Hosting concentration per type (2021)}
   \label{fig:conc-concentration-type}
\end{figure}

Figure \ref{fig:conc-concentration-type} shows the AS hosting country classification per TLD. Contrary to what we would expect, the US hosting industry is not dominant for most TLDs. The US hosting companies are popular in \dns{.ca} (same language), \dns{.co} (which is very popular outside Colombia), Guatemala's \dns{.gt} (possible due to geographical closeness to the US and services available in Spanish, which is Guatemala's official language) and Namibia's \dns{.na} (which has English as the official language).
%
%

Secondly, we see a solid local hosting industry in most European countries and Russia. The respective local companies are dominant in their local market. Interestingly, we see German-speaking countries in Europe (Austria's \dns{.at}, Switzerland's \dns{.ch}, and Liechtenstein' \dns{.li}) that use each other's countries hosting providers -- mostly German companies. Figure \ref{fig:conc-type-ipv4} shows the IPv4 breakdown. Austria hosts  42\% of its domains in German hosting companies, while Switzerland hosts 20\%. Liechtenstein has  60\% of its domains hosted by German or Swiss companies (30\% on each). The IPv6 hosting situation differs: we see a reduced presence in the local companies and a larger US companies based-hosting.

\begin{figure}[h!]
    \centering
    \begin{subfigure}[t]{0.5\textwidth}
        \centering
        \includegraphics[width=0.8\linewidth]{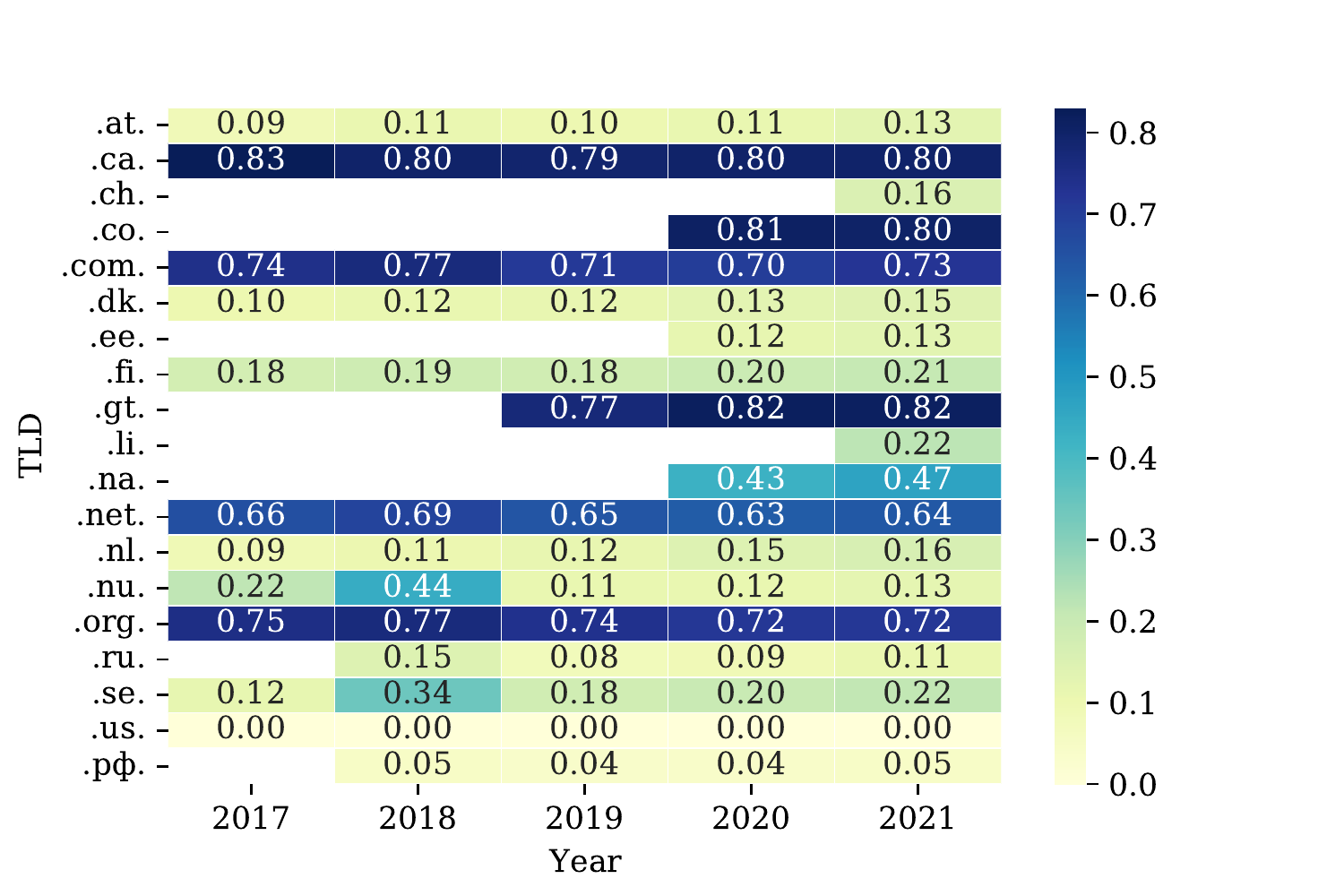}
        \caption{IPv4}
        \label{fig:v4-type-time}
    \end{subfigure}
    \begin{subfigure}[t]{0.5\textwidth}
        \centering
        \includegraphics[width=0.8\linewidth]{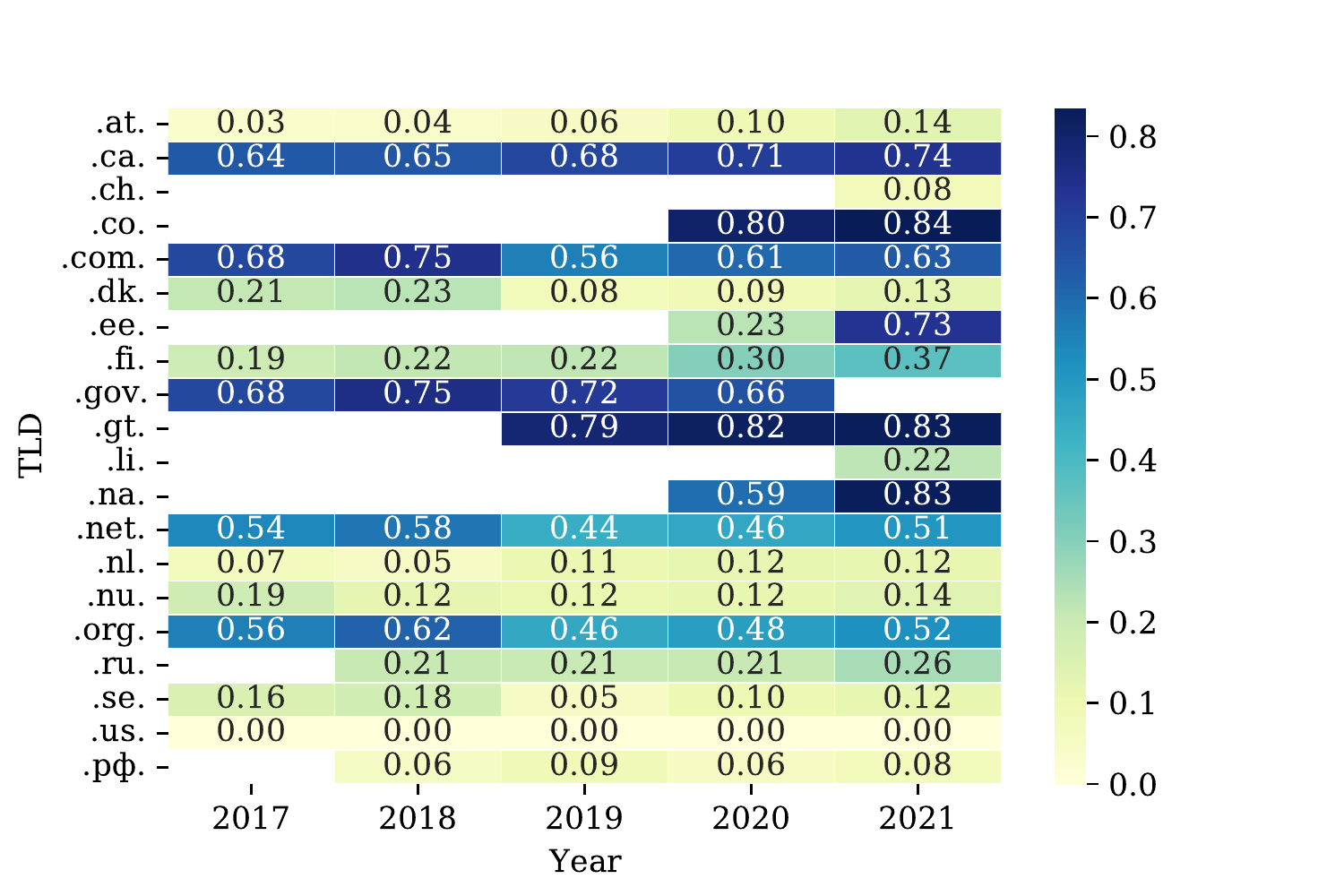}
        \caption{IPv6}
        \label{fig:v6-type-time}
    \end{subfigure}
    \caption{US hosting presence per TLD (IPv4)}
   \label{fig:type-time}
\end{figure}

\textit{US hosting presence growth:} Figure~\ref{fig:v4-type-time} shows the evolution of US hosting presence per TLD. For most TLDs we observe slow but continuous growth over time. Canada's \dns{.ca} has 80\% of its domains on US-based hosting companies for IPv4 (figure \ref{fig:v4-type-time}).

On the other extreme: Russia's ccTLDs (\dns{.ru} and $p\phi$) show a minor proportion of domains hosted by US companies (11\% and 4.7\%, respectively). Cloudflare, a US company, is among the few US cloud/content companies with a presence in Russia~\cite{ClouflareRU}, and hosts 6.3\% of the \dns{.ru} domain names. Russia is also known for having a controversial piece of legislation on privacy and local data storage~\cite{Moyakine2021a} and allegedly moving towards building isolated networks~\cite{Sherman2019a}. We can only speculate that these two factors contribute to the reduced American presence in Russia's hosting market.

In terms of recent growth, we see Sweden's \dns{.se} US-based hosting increased from 12\% to 22\% from 2017 to 2021. For IPv6, 73\% of Estonia's IPv6 hosting employs US-based companies (Figure \ref{fig:v6-type-time}). Both results pose challenges for EU policymakers, which have been concerned about Europe's digital sovereignty~\cite{Erlanger20a}.

\subsection{Hosting and domain name popularity}
\label{sec:host-popular}

The hosting industry has a significant variation in its service: from cheap, shared hosting for small businesses, to large to parking websites to anycast~\cite{RFC7094} widely distributed services~\cite{noroozian2015developing} \cite{tajalizadehkhoob2016apples}. 

Next, we investigate hosting concentration considering the \textit{popularity} of domain names as a proxy metric for the different segments within the hosting market. The idea is to determine the location of the companies hosting names on the Alexa Top list. It stands to reason that domains on the Alexa list are likely to use robust infrastructure.

To do that, we split each TLD dataset into two categories: if it is present or not on Alexa domain  lists~\cite{alexa} on the same day of the measurement.

\begin{figure}[h!]
    \centering
    \begin{subfigure}[t]{0.5\textwidth}
        \centering
        \includegraphics[width=0.8\linewidth]{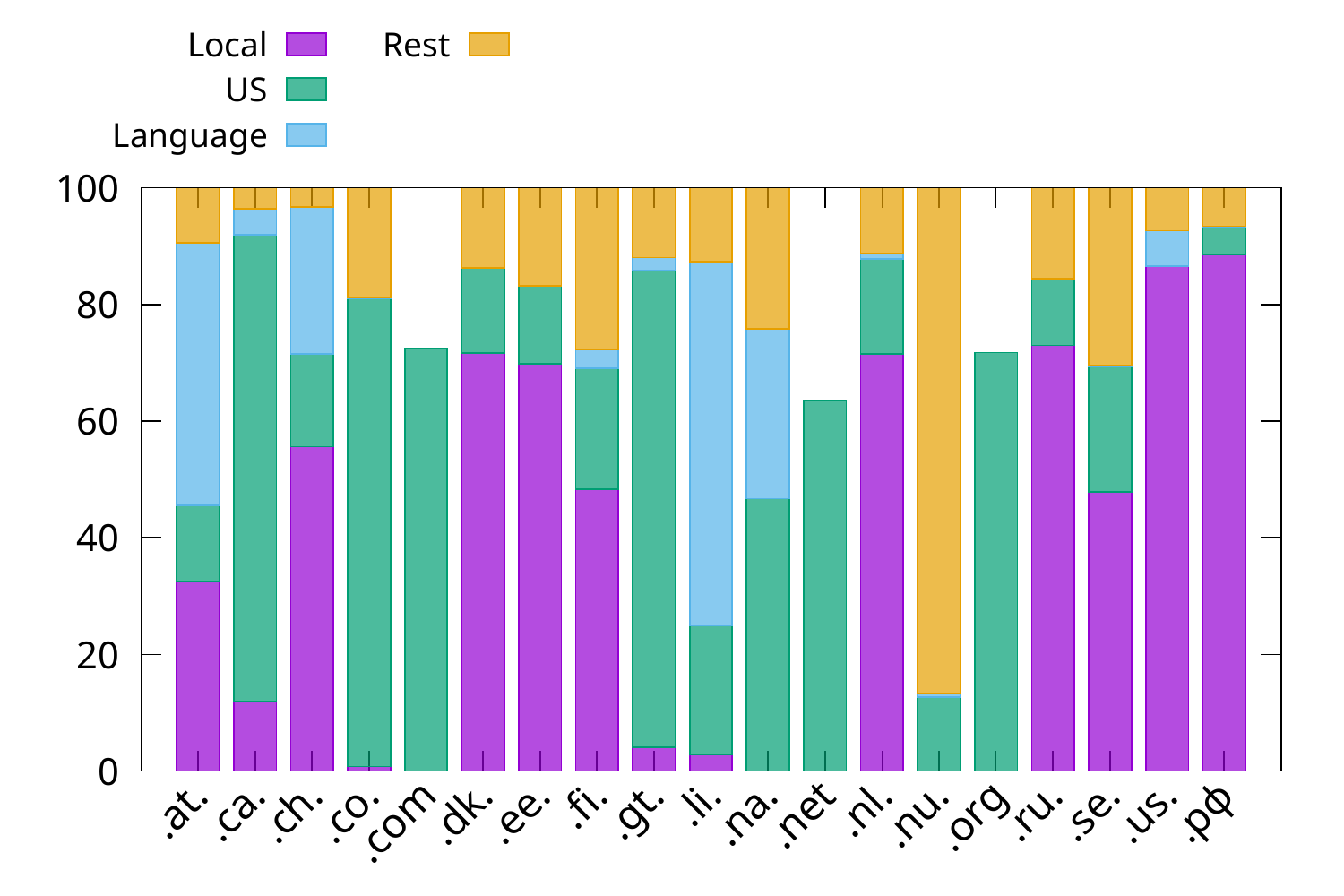}
        \caption{IPv4}
        \label{fig:v4-alexaFalse-2021}
    \end{subfigure}
    \begin{subfigure}[t]{0.5\textwidth}
        \centering
        \includegraphics[width=0.8\linewidth]{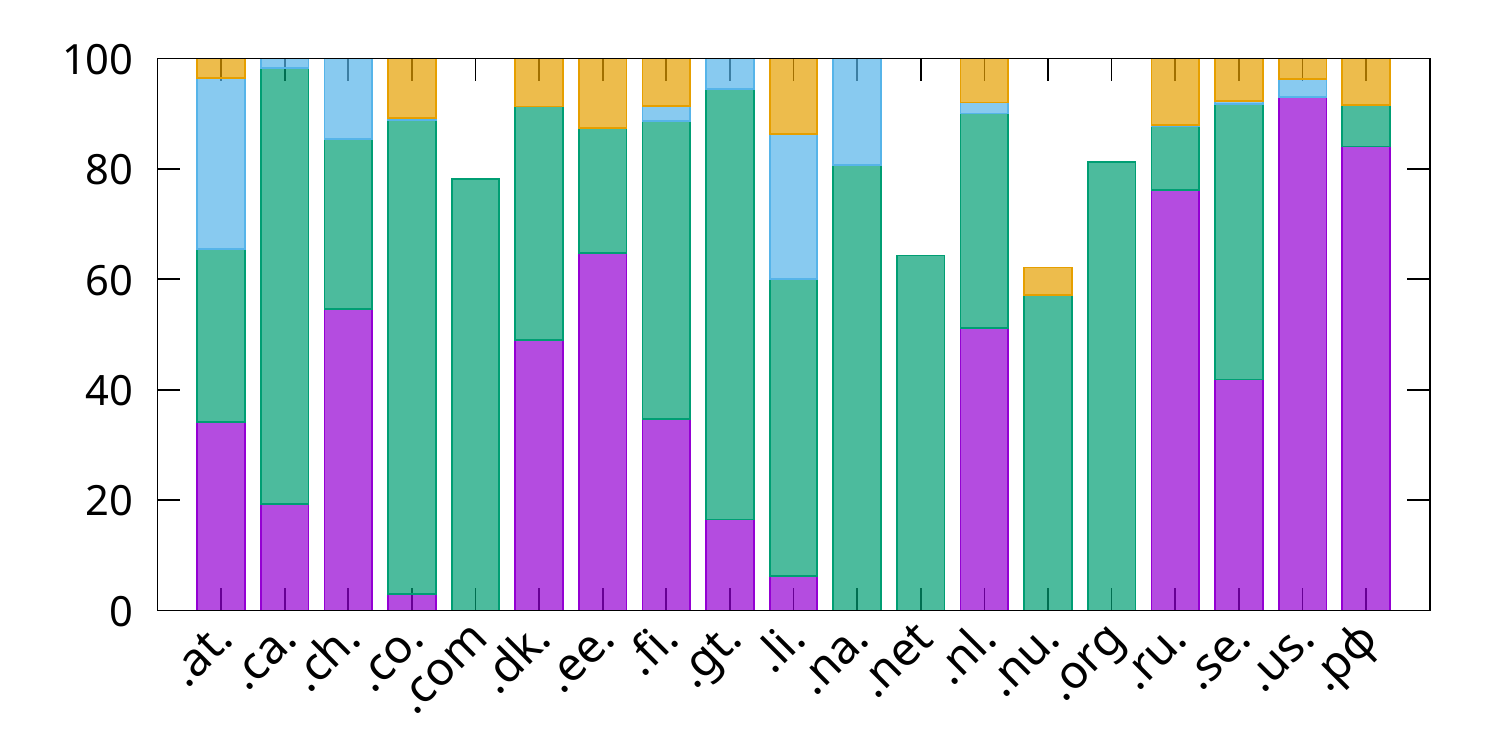}
        \caption{IPv6}
        \label{fig:v4-alexaTrue-2021}
    \end{subfigure}
    \caption{Hosting industry and domains popularity in 2021}
   \label{fig:alexa-time}
\end{figure}

Figure \ref{fig:alexa-time} shows the results for IPv4 in 2021. First, in \ref{fig:v4-alexaFalse-2021}, we see that the local providers dominates the hosting of domains outside the Alexa list. This, however, differs from what we see in \ref{fig:v4-alexaTrue-2021}, where we consider domains on the Alexa list. Here we see that most have more significant usage of US-based companies. It is interesting to see that Russian popular domains are mostly hosted by Russian companies. 

\begin{figure}[h!]
    \centering
    \begin{subfigure}[t]{0.5\textwidth}
        \centering
        \includegraphics[width=0.8\linewidth]{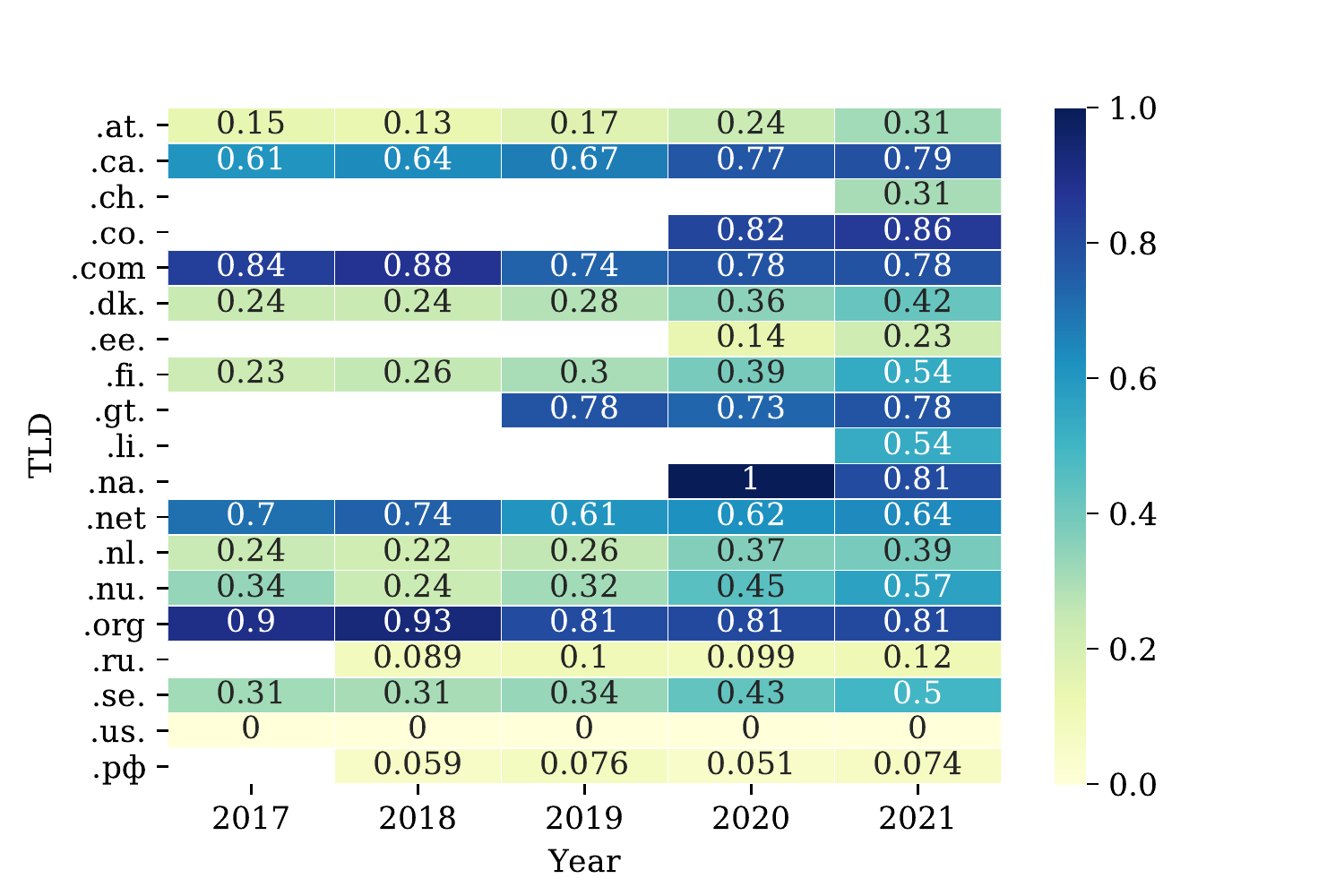}
        \caption{IPv4}
        \label{fig:alexav4-true-time}
    \end{subfigure}
    \begin{subfigure}[t]{0.5\textwidth}
        \centering
        \includegraphics[width=0.8\linewidth]{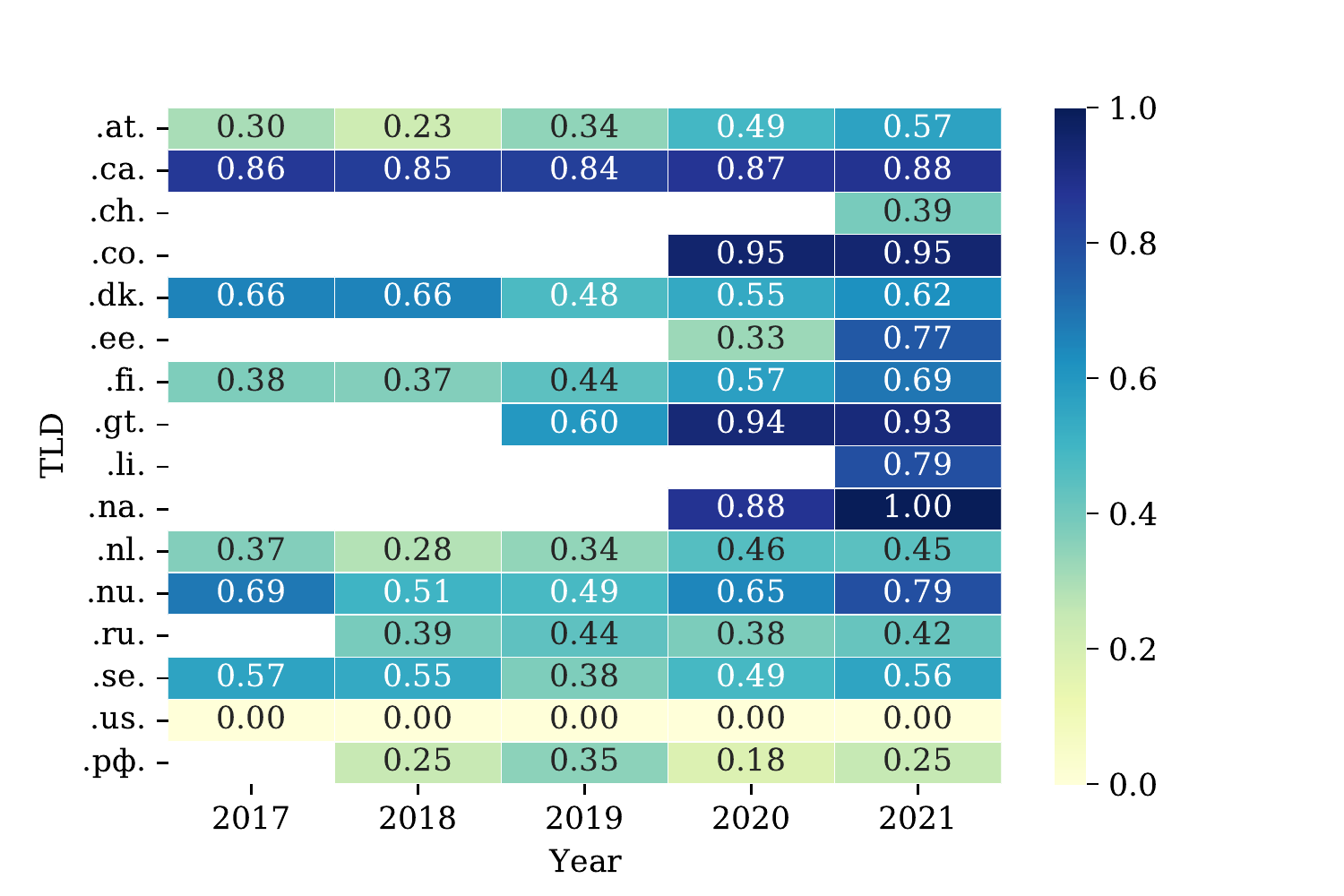}
        \caption{IPv6}
        \label{fig:alexav6-true-time}
    \end{subfigure}
    \caption{US hosting presence on Alexa domains over time}
   \label{fig:alexa-true-time}
\end{figure}

Figure \ref{fig:alexa-true-time} shows the evolution of usage of US hosting companies in Alexa domains for each TLD. Except for the Russian TLDs and Estonia's \dns{.ee}, we observe growth for the US hosting presence in Alexa ranked domains for all TLDs. These results suggest a trend among business of popular websites moving towards large companies.


%% file: src/conclusion.tex
\section{Conclusions and Future Work}
\label{sec:conclusion}

This work provides a new perspective on Internet centralization, complementing
previous works. By focusing on the hosting industry, we show how it is
heavily concentrated: 10 hosting providers account for most of the hosting for
all TLDs considered.  We show that European ccTLDs have a strong hosting industry.
However, US-based providers have been continuously conquering the market,
especially in the high end of it -- the popular domain names, which
poses challenges for the European Union's goals of digital sovereignty.  Russia,
on the other end of the spectrum, shows far less reliance on US-based
companies.

Future work will focus on addressing the limitations of
our hosting inferences described in Section III C. In particular, a systematic characterization of website content via web crawling will help further understand possible reasons behind centralization behaviors. Moreover, by using a classifier, we can treat parking domains as a particular case. 
Finally, coverage of (arbitrary)
fully qualified domain names can be increased by considering domain names learned from other data sources (e.g., Certificate Transparency Logs or Common Crawl Data)


%% file: src/acknowledgements.tex
\section{Acknowledgements}
\label{sec:acknowledgements}

We thank our reviewers for their constructive suggestions and feedback.
Funding for this work was provided in part by Capes PHD scholarship number 88887.480774/2020-00, the EU H2020 CONCORDIA
project (830927) and the NWO-DHS MADDVIPR project (628.001.031/FA8750-19-2-0004). This research used data from OpenINTEL,a project of the University of Twente, SURF, SIDN, and NLnet Labs.